\documentclass[11pt]{article}
\usepackage{latexsym,amsmath,amsthm,amssymb,amsfonts,amsbsy,wasysym,stmaryrd,hyperref
}
\usepackage[dvips]{graphicx,epsfig}
\usepackage{color}
\usepackage{url}

\usepackage{verbatim}

\parskip=\medskipamount

\csname @addtoreset\endcsname{equation}{section}

\usepackage{empheq}

\def\Real{{\mathbb R}}
\def\Comp{{\mathbb C}}

\def\1{1\hspace{-4pt}1}
\def\j1{\widetilde{1\hspace{-4pt}1}}

\def\bec{\begin{center}}
\def\ec{\end{center}}
\def\a{\alpha}
\def\ad{\dot{\a}}
\def\ua{{\underline \a}}

\def\b{\beta}
\def\bd{\dot{\beta}}

\def\e{\epsilon}

\def\k{\kappa}

\def\yb{{\bar y}}
\def\zb{{\bar z}}

\def\nn{\nonumber}
\newcommand{\eq}[1]{(\ref{#1})}

\def\ed{\end{document}}
\newcommand{\w}[1]{\\[0.#1cm]}
\def\be{\begin{equation}}
\def\ee{\end{equation}}
\def\bea{\begin{eqnarray}}
\def\eea{\end{eqnarray}}
\def\ba{\begin{array}}
\def\ea{\end{array}}

\def\ad{\dot\alpha}
\def\bd{\dot\beta}

\def\wt{\widetilde}

\definecolor{rougef}{rgb}{0.7,0,0}
\definecolor{vertf}{rgb}{0,0.6,0}
\definecolor{bleuf}{rgb}{0,0,0.9}

\def\bes{\begin{eqnarray}\begin{split}}
\def\ees{\end{split}\end{eqnarray}}

\begin{document}

\begin{center}


 {\Large\bf 4D Higher Spin Gravity with Dynamical Two-Form as a Frobenius-Chern-Simons Gauge Theory }


\vskip .5cm

{\bf Nicolas Boulanger$^1$, Ergin Sezgin\,$^2$ and Per Sundell\,$^3$}\\

\vskip .5cm

{\small{
{\em $^1$ \hskip -.1truecm
\textit{Physique Th\'eorique et Math\'ematique, 
Universit\'e de Mons -- UMONS\\ 
20 Place du Parc, B-7000 Mons, Belgium}
\vskip 1pt }

{email: {\tt nicolasboul@gmail.com}}
}}

\vskip 10pt

{\small{
{\em $^2$ \hskip -.1truecm George and Cynthia Woods Mitchell Institute for Fundamental
Physics and Astronomy \\ Texas A\& M University, College Station,
TX 77843, USA
\vskip 1pt }

{email: {\tt sezgin@tamu.edu}}
}}

\vskip 10pt

{\small{
{\em $^3$ \hskip -.1truecm
\textit{Departamento de Ciencias Fisicas, Universidad Andres Bello, Republica 220, Santiago de Chile}
\vskip 1pt }

{email: {\tt per.anders.sundell@gmail.com}}
}}

\vskip 0.3 cm

{\bf ABSTRACT}\\[3ex]
\end{center}

We provide an off-shell formulation of four-dimensional
higher spin gravity based on a covariant Hamiltonian 
action on an open nine-dimensional Poisson manifold whose
boundary consists of the direct product of spacetime and
a noncommutative twistor space of $S^2\times S^2$ topology.
The fundamental field is a superconnection consisting 
of even and odd differential forms valued in the odd and
even sectors of a 3-graded associative algebra given by the direct 
product of an eight-dimensional Frobenius 
algebra and a higher spin algebra extended by inner Klein operators.
The superconnection consists of two one-forms
gauging the one-sided actions of the 
higher spin algebra, two bi-fundamental real
forms given by the Weyl zero-form and a
new dynamical two-form, an additional set 
of forms providing a maximal duality extension, 
and, finally, the Lagrange multipliers required
for the covariant Hamiltonian action.
In a particular two-form background, the
model yields Vasiliev's recently proposed 
extended higher spin gravity equations,
whose interaction terms are accounted for 
by de Rham closed globally defined forms 
arising in the dynamical two-form. 


\pagebreak
\setcounter{page}{1}

\tableofcontents

\section{Introduction and summary}

  An outstanding problem in higher spin gravity 
\cite{Vasiliev:1990en,Vasiliev:1992av,Vasiliev:2003ev} is to find 
an action principle with desirable properties. Treating this 
problem as nonlinear completion of Fronsdal kinetic terms in a Noether procedure approach runs  into considerable technical difficulties. 
Indeed, in the metric-like \cite{Fronsdal:1978rb} and the related frame-like  
\cite{Vasiliev:1986td,Lopatin:1987hz,Vasiliev:2001wa} approaches, 
long term efforts --- see for example \cite{Fradkin:1986qy,Vasilev:2011xf}
and \cite{Buchbinder:2006eq,Metsaev:2006ui,Fotopoulos:2007yq,Zinoviev:2008ck,Boulanger:2008tg,Boulanger:2011qt,Joung:2012hz,Boulanger:2012dx,Joung:2013nma} 
in the case of $AdS$ background --- has so far led to primarily cubic interactions\footnote{
In flat spacetime and as far as quartic 
vertices are concerned, however, see 
\cite{Metsaev:1991mt,Taronna:2011kt,Dempster:2012vw} and references therein.}.
Beyond the cubic order, the fact that higher spin gravity
\footnote{We reserve the terminology of ``higher spin gravity'' to the higher spin extension of ordinary gravity by inclusion of integer spin massless fields represented by totally symmetric tensor fields, while ``higher spin theory'' will refer to a larger classes of theories in which mixed symmetric higher spin fields and/or massive higher spin fields arise as well.}
has a mass scale 
set by the bare cosmological constant while nonabelian higher spin symmetries  require higher derivative vertices, lead to intractable abelian vertices built from curvatures and their higher derivatives;  see \cite{Boulanger:2008tg} and the review \cite{Bekaert:2010hw}.

The Noether procedure approach does not exploit the fact that 
Vasiliev's equations \cite{Vasiliev:1990en,Vasiliev:1992av,Vasiliev:2003ev} 
provide a fully non-linear description of higher spin  gravity on-shell. Furthermore, 
it is background dependent procedure since it  is based on perturbation around $AdS_4$. 
Both drawbacks can be avoided by considering covariant Hamiltonian actions from which the 
background independent full Vasiliev equations follow. 
These equations are Cartan integrable systems of differential 
forms on special noncommutative manifolds taking their values in 
associative higher spin algebras.
Treating these forms as the fundamental fields following the AKSZ approach
\cite{Alexandrov:1995kv}, one is led to a path integral formulation \cite{Boulanger:2012bj} 
based on covariant Hamiltonian actions \cite{Boulanger:2011dd} on 
noncommutative manifolds with boundaries.
The importance of the boundaries, which are absent
in the related proposals \cite{Vasiliev:1988sa,Doroud:2011xs},
is that they facilitate the deformation of the 
bulk action by boundary terms \cite{Sezgin:2011hq}
that contribute to the action but not its variation on-shell. 
The resulting higher spin amplitudes reproduce desired 
holographic correlation functions 
\cite{Colombo:2010fu,Colombo:2012jx,Didenko:2012tv},
which are suggestive of an underlying topological open string
\cite{Engquist:2005yt,Arias:2015wha,Bekaert:2015tva}. 
This framework connects holographic duals to topological 
theories in two higher dimensions via intermediate boundary states 
described by topological invariants on-shell \cite{Sezgin:2011hq},
without referring to any Fronsdal kinetic terms. However,
the presence of a large number of free parameters impede the 
predictive power of the model.

  The purpose of this paper is to modify Vasiliev's theory without 
altering its local degrees of freedom\footnote{By local 
degrees of freedom, we mean the representation space containing the
integration constant for Vasiliev's master zero-form $B\,$.} 
but nonetheless introducing more gauge symmetries,
hence leading to more predictive models.
We shall focus on four-dimensional higher spin gravities, 
mainly for the sake of simplicity. The resulting model based 
on an action will be referred to as Frobenius-Chern-Simons (FCS) gauge theory
for reasons that wil be explained below.
In the remainder of this section, we shall expand on the points mentioned 
briefly above  with regard to the nature of Vasiliev's equations, the 
existing action proposals, the role of on-shell topological invariants, and 
we shall summarize the key results of this paper which provides an FCS model
is proposed.

  Vasiliev's original formulation \cite{Vasiliev:1990en,Vasiliev:1992av}
takes place on the direct product ${\cal M}_4\times
{\cal Z}_4$ of a commutative manifold ${\cal M}_4$
(possibly with boundaries) and a noncommutative twistor
space ${\cal Z}_4\,$.
In maximally symmetric backgrounds, infinite towers of Fronsdal 
tensors on ${\cal M}_4$, including matter fields, are packaged 
together with auxiliary fields into a one-form $W$ and a zero-form 
$B$ on ${\cal M}_4\times{\cal Z}_4$, valued in an oscillator algebra 
extended by outer Klein operators.
The emerging structure is an associative differential 
algebra consisting of forms on a noncommutative base manifold 
that are valued in an associative $\star$ product algebra. 
The curvature two-form $F=dW+W\star W$ and a covariant
derivative $DB=dB+W\star B-B\star W$ obeying the
Bianchi identities $DF=0$ and $D^2 B=[F,B]_\star\,$.
While the Bianchi identities permit setting $F$ and 
$DB$ to zero, the Weyl curvatures of the 
original Fronsdal fields are mapped to a 
deformation of $F$ proportional to a special
{\it closed and central holomorphic two-form $J$ on 
${\cal Z}_4$}, \emph{viz.}
\be
dJ=0\ ,\qquad [J,B]_\star=0=[J,W]_\star\ ,
\ee
and its hermitian conjugate $\overline J$, 
valued in an extension of the aforementioned 
oscillator algebra by inner as well as outer Klein operators.
Prior to duality extension and further extensions
by closed and central elements \cite{Boulanger:2011dd,
Vasiliev:2015mka}, that we shall attend below, Vasiliev's 
original equations read
\be
dW+W\star W+{\cal V}(B)\star J-\overline{\cal V}(B)\star \overline J=0\ ,\qquad
dB+W\star B-B\star W=0\ ,
\label{VE}
\ee
where the star function ${\cal V}(B)$ represents an interaction
ambiguity, which is defined on-shell modulo field redefinitions.
Working perturbatively in $B$, one may take ${\cal V}=B\star \exp_\star (i\Theta(B))$
where $\Theta$ is a real star function,
which can be fixed by demanding the existence
of a consistent truncation to one of the two parity
invariant minimal bosonic models, implying
${\cal V}=B$ and ${\cal V}=iB$ in the cases of the Type A
and Type B models, respectively \cite{Sezgin:2003pt}.

  In order to integrate the equations of motion, which form
a locally defined Cartan integrable system, into a well 
defined action, one needs to introduce 
additional geometric structures.
In \cite{Sezgin:2011hq}, higher spin geometries were
introduced by making use of structure groups and
the universal Cartan integrability of Vasiliev's equations,
which permits the introduction of extra coordinates 
without affecting the number of local degrees of freedom
\cite{Engquist:2002gy}. 
In the resulting higher spin geometries, which in general
are formulated on higher dimensional extensions of 
${\cal M}_4$, a number of globally defined higher
spin invariants can be constructed \cite{Sezgin:2011hq}.
A subset of these, namely those that arise from topological
invariants, are suitable building blocks for an effective action, 
as they can be added to the aforementioned covariant Hamiltonian 
bulk action, which vanishes on-shell for the simplest types of boundary
conditions, without affecting the equations of motion while producing
nontrivial contributions on-shell.
In principle, when evaluated on asymptotically $AdS_4$ solutions, the 
resulting free energy will be given as a function of masses and other 
charges accessible to an infrared asymptotic observer, providing a
fundamental thermodynamics relation.

  The simplest invariants, found in \cite{Sezgin:2005pv,Engquist:2005yt}, 
are integrals over ${\cal Z}_4$ of traces over the oscillator algebra
evaluated at a base point $p_0\in {\cal M}_4$, 
hence referred to as zero-form charges, \emph{viz.}\footnote{
Generalizations obtained by inserting open 
Wilson loops on ${\cal Z}_4$ and further (odd)
powers of $B$ can be found in \cite{Colombo:2010fu,Sezgin:2011hq}.}
\be {\cal I}_{2n}= \int_{{\cal Z}_4}
{\rm Tr}  \left.(k\star \bar k\star
B^{\star (2n)} \star J\star \overline J )\right\vert_{p_0}\ ,\label{I2n}
\ee
using a notation to be described in Section 3.
These functionals
can also be seen to exhibit cluster decomposition
properties \cite{Iazeolla:2011cb} that turn them into
natural building blocks for higher spin amplitudes
\cite{Colombo:2010fu}, hence referred to as quasi-amplitudes,
examined in more detail in \cite{Colombo:2012jx,Didenko:2012tv}.
In an attempt to incorporate these building blocks into an effective
action, it was proposed in \cite{Boulanger:2011dd,Boulanger:2012bj} to apply the 
AKSZ formalism \cite{Alexandrov:1995kv}\footnote{
The AKSZ formalism, which is an adaptation of the
manifestly covariant Batalin--Vilkovisky  quantization
\cite{Batalin:1981jr,Batalin:1984jr} to topological
field theories, was originally applied to topological
open strings and $p$-branes
\cite{Cattaneo:2001ys,Ikeda:2001fq,Hofman:2002jz,Roytenberg:2006qz};
for related topics, see \cite{Barnich:2009jy,Alkalaev:2013hta}.}
to Vasiliev's equations by introducing an auxiliary dimension
and considering an action on ${\cal M}_9 = {\cal X}_5 \times {\cal Z}_4$
where $\partial{\cal Z}_4=\emptyset$ and ${\cal X}_5$ is
an open manifold whose boundary contains ${\cal M}_4$
as a subregion.
The action, which is of a covariant Hamiltonian form,
reproduces Eq. \eq{VE} on $\partial{\cal M}_9$ but 
vanishes on-shell.

  In order to generate an on-shell action,
it was proposed in \cite{Sezgin:2011hq} to 
add boundary terms (on $\partial{\cal M}_9$), 
referred to as topological vertex operators, to
the bulk action (on ${\cal M}_9$).
Their total variations vanish on-shell, and 
hence they reduce on-shell to invariant functionals.
In particular, it was shown how to incorporate some 
quasi-amplitudes, still infinite in number, into the 
effective action.
However, as each such building block appears with its own free
parameter, the resulting model contains a large number of
undetermined coupling constants. This shortcoming is due to the fact that $B$ has form degree
zero and transforms in the adjoint representation of the higher spin algebra
\footnote{Certain on-shell higher spin invariant functional have been proposed as generating
functional for correlators within the context of holography \cite{Vasiliev:2015mka}. 
We shall comment further on this approach in Section 5.2, where we shall also compare it 
to the one in which we add suitable topological invariants to the covariant Hamiltonian action.}. 

To improve predictability, we propose to modify the theory such that $B$ transforms in a 
bifundamental representation of an enlarged symmetry algebra, \emph{viz.}
\be 
\delta B=-\e\star B+B\star \tilde \e\ .
\ee
The definition of a covariant derivative $DB$ requires the introduction of an additional 
one-form master gauge field ${\wt A}$ such that $DB=dB+A\star B-B\star 
\widetilde A$. 
The form of the resulting Bianchi identity, \emph{viz.} $D^2B=F\star B-B\star \widetilde F$ 
where $F=dA+A\star A$ and $\wt F=d\wt A+\wt A\star \wt A$, and the fact
that $F$ and $\wt F$ are algebraically independent, can be dealt with by replacing the 
rigid two-forms $J$ and $\overline J$ by a dynamical two-form master field $\wt B$ 
transforming as
\be 
\delta \widetilde B=-\tilde\e\star\widetilde B+\widetilde B\star\e\ ,
\ee
after which one may consider the following Cartan integrable system:
\begin{eqnarray}
& dA + A\star A-B\star \widetilde B = 0\ ,\qquad d\wt A + \wt A \star \wt A - \wt B\star B=0 \ ,&
\nonumber \\
& dB+A\star B - B \star \wt A = 0\ ,\qquad d \wt B + \wt A \star \wt B - \wt B \star A = 0 \ .&
\label{master}
\end{eqnarray}
The simplest invariant are now given by integrals of traces of $(B\star \widetilde B)^{\star n}$ (or $(\widetilde B\star B)^{\star n}$), which provided a finite 
number of building blocks for an effective on-shell action (provided that 
$\partial {\cal M}_4\times{\cal Z}_4$ contains a finite number 
of de Rham cocycles); for example, on $M_9$ we have $n\leqslant 4$. 
The interaction ambiguity ${\cal V}(B)$ can be accounted for by demanding that 
$\widetilde B$ can acquire vacuum expectation
values of the form\footnote{These backgrounds can be enriched
by de Rham cohomology elements from ${\cal X}_4\,$.}
$ \widetilde B= f(B,J,\overline{J})$. In order for integrals over ${\cal Z}_4$
to be well-defined and for $(J,\bar J)$ to be globally
defined on ${\cal Z}_4$, we shall choose  ${\cal Z}_4=S^2\times S^2$.
The original Vasiliev system then emerges as the 
consistent truncation $A=\wt A\equiv W$ combined with the ansatz
$ \widetilde B= f(B,J,\overline{J})$.

A key property of Eqs. \eqref{master} is that they can be expressed 
as a vanishing curvature condition by employing an eight-dimensional, 
3-graded Frobenius algebra 
\be {\cal F}={\cal F}^{(-1)}\oplus{\cal F}^{(0)}\oplus {\cal F}^{(+1)}\ ,
\qquad (e_{ij}, h\, e_{ij})\in {\cal F}^{(i-j)}\ ,\ee
where $e_{ij}$ is the $2\times 2$ matrix whose only non-vanishing entry
is a $1$ at the $i$th row and $j$th column, and $h$ is a Klein 
element satisfying $[h,e_{11}]=0=[h,e_{22}]\,,\;\{h,e_{12}\}=0 = \{h,e_{12}\}\,$.
Assembling the master fields into
\be
X = \sum _{i,j} X^{ij} e_{ij} = \begin{pmatrix}
A & B \\ \widetilde B & \widetilde A 
\end{pmatrix}\ ,
\ee
the equations \eq{master} can be written as the flatness conditions
\be F^X := dX +hXh\star X=0\ .\ee
As in \cite{Boulanger:2011dd}, we then introduce the
Lagrange multiplier master field
\be
P = \sum _{i,j} P^{ij} e_{ij} = \begin{pmatrix}
V & U \\   \widetilde U & \widetilde V
\end{pmatrix}
\ee
and duality extend\footnote{The precise definition
of duality extension \emph{on-shell}, which was made in Appendix
D of \cite{Boulanger:2011dd}, involves the \emph{simultaneous}
introduction of new dynamical higher forms and an algebra of
closed and graded central elements, so as to glue nontrivially the
original forms to the new forms, as opposed to the more
restrictive notion of \emph{projective extension}
discussed in \cite{Vasiliev:2007yc}, which only introduces higher
forms but no graded central and closed elements.
} the field content off-shell, to obtain forms with degrees
\begin{eqnarray}  
& {\rm deg} (B,A,\widetilde A,\widetilde B)\in
\left\{(2n,1+2n,1+2n,2+2n)\right\}_{n=0,1,2,3}\ ,&
\nonumber \\
& {\rm deg} 
(\widetilde U,V,\widetilde V,U)=\left\{(8-2n,7-2n,7-2n,6-2n)
\right\}_{n=0,1,2,3}\ . &
\label{contentDE}
\end{eqnarray}
We proceed by defining a superconnection and superdifferential \cite{Quillen198589} as
\be 
Z = hX+ P~\in~ \Omega({\cal M}_9)\otimes {\cal A}\ ,\qquad q=hd\ ,
\ee
respectively, which are thus objects with odd superdegree 
given by form degree plus Frobenius 3-degree. The algebra ${\cal A}$, which 
will be referred to as the higher spin algebra, contains the Frobenius algebra 
as a factor and it will be defined below. Employing these ingredients, we construct the action
\footnote{The action does not truncate to
the one constructed in \cite{Boulanger:2011dd} under $ \widetilde B
= f(B,J,\overline{J})$ and $A=\tilde A\equiv W\,$.
In fact, it can be recuperated from the FCS 
action by rescaling the fields and taking a limit but  
this sends to zero the nontrivial Poisson $3\,$-vector field, thereby
trivialising the action from the point-of-view of the category of 
covariant Hamiltonian systems.
}
\bea
S  &=&  \int_{{\cal M}_9} {\rm Tr}_{{\cal A}}
\left(\tfrac{1}{2}  \,Z\star q Z+\tfrac{1}{3}\, Z\star Z\star Z\right)
-\frac{1}{4}\, \int_{\partial{\cal M}_9} {\rm Tr}_{{\cal A}} \,[h\pi_h(Z) \star Z]
\nn\w2
 &=&
\int_{{\cal M}_9} {\rm Tr}_{{\cal A}} \, \left(P \star F^X
+\tfrac13 \,P\star P\star P\right) \ ,
\eea
where $\pi_h$ is the automorphism sending $h$ to $-h\,$.
The boundary term is essential for a globally well-defined
action functional, and under the assumption that $P$ forms a section obeying
$P\vert_{\partial{{\cal M}_9}}=0\,$, 
the equations of motion reduce to the duality extended version 
of \eq{master} at $\partial{\cal M}_9\,$. 
We shall refer to the above natural generalization of Chern--Simons
theory as Frobenius--Chern--Simons (FCS) theory.
Among its properties, we stress the following:
\begin{itemize}

\item[i)] The moduli of $\wt B$ contain closed and central
elements on ${\cal X}_4\times {\cal Z}_4\,$, which combine
with integration constants from $B_{[0]}$ into deformations 
of the curvatures on-shell.
The cycles in degree two cause deformations of the 
Poisson structure on ${\cal Z}_4\,$.
The cycles in higher degrees are required for the 
forms in degrees greater than two, introduced by the duality 
extension, to receive nontrivial corrections on shell;

\item[ii)] The model contains a particular two-form background
around which the expansion of the equations of motion yields  
Vasiliev's equations upon fixing a gauge for $\wt B$ and $\wt A -A$.
In particular, standard Lorentz covariance arises within this
phase of the theory.

\item[iii)] The $P^{\star 3}$ term in the action, which 
requires the duality extension, may lead to radiative 
corrections to boundary correlation functions from bulk 
Feynman diagrams.
They may also lead to novel contributions to the classical 
action when ${\cal M}_9$ has multiply connected boundaries.
\item[iv)] The formulation of the model admits a natural 
extension to general noncommutative manifolds, akin to that expected 
from an underlying topological open string \cite{Engquist:2005yt,
Arias:2015wha,Bonezzi:2015lfa}.

\end{itemize}

  This paper is organized as follows: 
In Section \ref{sec:BaseManifold}, we give 
the basic properties of the noncommutative space
${\cal Z}_4 = S^2\times S^2\,$.
The Frobenius algebra ${\cal F}\,$, the higher spin algebra 
${\cal W}$ its trace operation are spelled out 
in Section \ref{sec:HigherSpinAlgebra}.
We then turn in Section \ref{sec:FCSmodel} 
to the action, equations of motion,
global formulation, including boundary conditions, and various
projections of the model, including the minimal bosonic one.
In Section \ref{sec:Vasi}, we compare a certain truncation of 
our model, that activates closed and graded central elements, 
to Vasiliev's duality extended model, which includes Lagrangian 
forms.
We will stress the difference between the Lagrangian forms 
and the aforementioned on-shell invariants for which we
propose a physical interpretation within an effective action
based on a path integral approach.
In Section \ref{sec:Boundary}, we examine the perturbative 
fluctuations on ${\cal M}_4\times {\cal Z}_4$ and show that 
the local degrees of freedom of the FCS model
are the same as those of original Vasiliev system, while
the fluctuations in $A-\wt A$ and $\wt B$ may introduce new 
topological degrees of freedom.
We also show how the boundary conditions on ${\cal Z}_4$ 
lead to globally defined Killing symmetries in agreement with those 
of Vasiliev's original model.
In Section \ref{sec:Conclusions} we comment further on our results and 
point out open problems and future directions. 

%

\section{Base manifold}
\label{sec:BaseManifold}

  The model will be formulated in terms of differential forms on the direct product space
\be {\cal M}_{9}= {\cal X}_{5}\times {\cal Z}_4\ ,\label{calBtop}\ee
where ${\cal X}_{5}$ is a five-dimensional commutative
manifold with boundary
\be {\cal X}_{4}=\partial {\cal X}_{5}\ ,\ee
containing the original spacetime manifold ${\cal M}_4$ as a
possibly open subset,
and ${\cal Z}_4$ is a four-dimensional
noncommutative space without boundary.
Thus,
\be
\partial{\cal M}_9={\cal X}_4\times {\cal Z}_4\ .
\ee
We shall make the simple choice 
${\cal X}_5={\cal X}_4\times [0,\infty[\,$. 
An interesting alternative would be 
${\cal X}_5={\cal X}_4\times [0,1]\,$, 
which would yield two copies of the action on the two 
boundaries of the cylinder, and there may be interesting 
interactions interpolating between them.
We shall not consider this possibility further here. 


  The topology of ${\cal Z}_4$ may be chosen in a variety 
ways with nontrivial and interesting consequences that 
we hope to treat elsewhere.
In what follows, we shall obtain ${\cal Z}_4$ from the 
standard noncommutative $\Comp^4$ by choosing a real 
form and a compatible convolution formula for the star 
product and then adding points at infinity to create a 
compact noncommutative space that can be used to define 
a (graded cyclic) trace operation.
Moreover, we shall demand ${\cal Z}_4$ to be closed,
to avoid boundary terms, and that its structure admits 
a certain closed two-form and a global $SL(2;\Comp)$ 
symmetry, in order to make contact with Vasiliev's theory.

  To this end, we introduce canonical coordinates
$(z^\a,\bar z^{\ad})$ ($\a,\ad=1,2$) and anti-commuting
differentials $(dz^\a,d\bar z^{\ad})$ on $\Comp^4$.
We then consider a formally defined associative star product algebra
given by the space $\Omega(\Comp^4)$ of differential forms
equipped with two associative composition rules, namely the
standard graded commutative wedge product rule, denoted by juxtaposition,
and the graded noncommutative rule
\be f\star g= f \exp\left( - i(\overleftarrow {\partial}^{\a}\overrightarrow{\partial_{\a}}
+\overleftarrow {\bar\partial}^{\ad}\overrightarrow{\bar\partial}_{\ad})\right)\, g\ ,
\label{starZ}\ee
which is compatible with the ordinary differential, \emph{viz.}
\be d(f\star g)=df\star g+(-1)^{{\rm deg}(f)}f\star dg\ .\ee
The star product is thus the representation using Weyl ordering
symbols of the associative algebra of composite operators
built from anti-commuting line elements and noncommutative
coordinates with canonical commutation rules
\be 
[z^\a,z^\b]_\star=-2i\e^{\a\b}\ ,\qquad [z^\a, z^{\ad}]_\star=0\ ,
\qquad [\bar z^{\ad}, \bar z^{\bd}]_\star=-2i\e^{\ad\bd}\ .
\ee

  In order impose reality conditions on the FCS model
and project it further to minimal models, one needs
to employ linear and anti-linear anti-automorphisms,
denoted by $\tau$ and $\dagger$, respectively, that are compatible with 
the basic algebraic structures, \emph{i.e.}
\be  
(f\star g)^\dagger=(-1)^{{\rm deg}(f){\rm deg}(g)}g^\dagger\star f^\dagger\ ,\qquad
(df)^\dagger=d(f^\dagger)\ .\label{daggercompatibility}\ee
\be 
\tau(f\star g)=(-1)^{{\rm deg}(f){\rm deg}(g)}\tau(g)\star \tau(f)\ ,\quad 
\tau(df)=d(\tau(f))\ .\label{taucompatibility} 
\ee
In models with four-dimensional Lorentz symmetry,
it is natural to select real forms on the real slice\footnote{
A graded differential star product algebra
$\Omega$ with a compatible hermitian conjugation
can be decomposed into real and imaginary parts.
The imaginary subspace remains closed under 
graded skew symmetric star product commutators,
and hence defines a real graded Lie 
subalgebra of $\Omega$.
Real forms of $\Omega$ that are preserved by the star product 
instead require
anti-linear automorphisms $\ast$, \emph{a.k.a.} star
maps, obeying $(f\star g)^\ast
= f^\ast \star g^\ast$ and $(df)^\ast=d(f^\ast)$.
These can be obtained as $f^\ast=\tau(f^\dagger)$. }
\be {\cal R}^{\Comp}_4=\left\{ (z^\a,\bar z^{\ad})\,:\ (z^\a)^\dagger=-\bar z^{\ad}\ ,\quad
(\bar z^{\ad})^\dagger=-z^\a\,\right\}\, \cong \, \Comp^2 \times \overline {\Comp^2}\,\subset\, \Comp^4\ ,\label{realZ}\ee
on which $z^\a$ is thus a complex doublet, and we note that $(f^\dagger)^\dagger=f\,$. 
We shall also use
\be 
\tau(z^\a,\bar z^{\ad}) = (-iz^\a,-i\bar z^{\ad})\ .\label{tau}
\ee

  In order to include Gaussian elements and distributions, it is useful 
to first introduce auxiliary integral representations of the star product \eq{starZ}.
As $[z^\a,\bar z^{\ad}]_\star=0$, there are two
natural convolution formulae, depending on whether
the auxiliary variables are complex or real doublets.
We shall choose the latter, \emph{viz.}
\be f\star g= \int_{{\cal R}^{\Real}_4}  \frac{d^2\xi d^2\tilde\xi}{(2\pi)^2} 
\int_{{\cal R}^\Real_4} \frac{d^2\eta d^2\tilde\eta}{(2\pi)^2}
e^{i(\eta^\a \xi_\a+\tilde\eta^{\ad}\tilde\xi_{\ad})}
f(z+\xi,\bar z+\tilde\xi; dz,d\bar z) g(z-\eta,\bar z-\tilde \eta;dz,d\bar z) ,
\label{starZI}\ee
where the integration domain\footnote{Alternatively, one
can define a star product 
using ${\cal R}_\Comp$ as integration domain, but as we shall see the construction
of the vacuum expectation value for the two form leading
to higher spin gravity requires the choice in \eq{starZI}.
}
\be
{\cal R}^{\Real}_4 = \left\{(\xi^\a,\tilde\xi^{\ad}): \ \xi^\a,\tilde\xi^{\ad} \in \Real^2\right\}\cong \Real^2\times \Real^2\ .
\label{calRR}
\ee
We note that even though different
real slices are used in \eq{realZ} and \eq{starZI},
the compatibility condition \eq{daggercompatibility}
still holds.

  In order to include Gaussian functions, we rely on complex-analytic 
continuation in the eigenvalues of the bilinear forms involved.
This requires the further inclusion of delta-function distributions
of the form $\delta^2(M^\a_\b z^\a)$ and $\delta^2(\bar M^{\ad}_{\bd}\bar z^{\ad})$,
with complex matrices, for which we need to use the prescription 
\footnote{
For an application of this distribution in
the context of exact solutions, see \cite{Iazeolla:2011cb}.}
\be \delta^2(M^\a_\b z^\b)=
(\det M)^{-1} \delta^2(z^\a)\ ,\label{deltaM}\ee
without any absolute value, after which $\delta^2(z^\a)$ is treated
by first rotating (the constants) $(z^\a,\bar z^{\ad})$ from 
${\cal R}^\Comp_4$ to ${\cal R}^\Real_4$, then performing auxiliary integrals,
and finally rotating $(z^\a,\bar z^{\ad})$ back to ${\cal R}^\Comp_4$.
More explicitly, following this prescription
\bea
\delta^2(M^\a_\b z^\b)\star f(z^\a)&=&(\det M)^{-1}
\int_{{\cal R}^{\Real}_4}  \frac{d^2\xi d^2\tilde\xi}{(2\pi)^2} 
\int_{{\cal R}^\Real_4} \frac{d^2\eta d^2\tilde\eta}{(2\pi)^2}
e^{i(\eta^\a \xi_\a+\tilde\eta^{\ad}\tilde\xi_{\ad})}
\delta^2(z+\xi)f(z-\eta)
\nn\w2
 &=&(\det M)^{-1} 
\int_{\Real^2}\frac{d^2\xi }{(2\pi)^2} 
e^{i z^\a \eta_\a}f(z-\eta)\ ,
\eea
where $z^\a$ is thus treated as a real doublet in the intermediate expression,
and, likewise, 
\bea
f(z^\a)\star \delta^2(M^\a_\b z^\b) &=& (\det M)^{-1}
\int_{{\cal R}^{\Real}_4}  \frac{d^2\xi d^2\tilde\xi}{(2\pi)^2} 
\int_{{\cal R}^\Real_4} \frac{d^2\eta d^2\tilde\eta}{(2\pi)^2}
e^{i(\eta^\a \xi_\a+\tilde\eta^{\ad}\tilde\xi_{\ad})}
f(z+\xi)\delta^2(z^\a-\eta^\a)
\nn\w2
&=& (\det M)^{-1} 
\int_{\Real^2}\frac{d^2\xi }{(2\pi)^2}  e^{iz^\a \xi_\a}f(z+\xi)\ .
\eea
Thus, the delta functions are required to be treated as 
complex analytic distributions rather than real analytic ones.

  The need for the prescription \eq{deltaM} can be seen by examining star
products involving the Gaussian function $e^{\lambda z^1 z^2}$ with $\lambda\in \Comp$,
\emph{viz.}
\be e^{\lambda z^1 z^2}\star f(z)=\int_{{\cal R}^\Real_4} 
\frac{d^2\xi d^2\eta}{(2\pi)^2}
e^{i\eta^\a \xi_\a+\lambda(z^1+\xi^1)(z^2+\xi^2)}f(z-\eta)\ ,\ee
whose left-hand side can be computed directly using \eq{starZ},
which leads to a formula that is a complex analytic function of
$\lambda$, while the right-hand side can be computed in two ways,
depending on whether the integral is performed by integrating
$\xi^1$ and $\xi^2$ one at a time using \eq{deltaM} or by 
Gaussian integration using analytical continuation on $\lambda$.
To illustrate this point, we may focus on the integral 
\be I(\lambda)=\int_{\Real^2} d\xi^1 d\xi^2 e^{\lambda \xi^1 \xi^2}\ .\ee{%
To bring it to Gaussian form, we change variables to $u=\frac12(\xi^1+\xi^2)$
and $v=\frac12(\xi^1-\xi^2)$, which brings the integral to 
Fresnel?s form if $\lambda$ is purely imaginary.
Writing $\lambda=i\mu$ with $\mu$ a non-zero real number, we find
\be I(\lambda)= 8 \int_0^\infty du e^{i\mu u^2} \int_0^\infty dv e^{-i\mu v^2}=
\frac{2\pi}{\mu}=\frac{2\pi i}{\lambda}\ .
\label{fresnel}\ee
On the other hand, integrating the $\xi^1$ and $\xi^2$ variables 
one at a time using $e^{\lambda\xi^1\xi^2}=e^{i(-i\lambda\xi^1)\xi^2}$
and the prescription \eq{deltaM} yields 
\be I(\lambda)=\int_{-\infty}^\infty d\xi^1 \int_{-\infty}^\infty d\xi^2 
e^{i(-i\lambda\xi^1)\xi^2}
= 2\pi\int_{-\infty}^\infty d\xi^1 \delta(-i\lambda\xi^1)= \frac{2\pi}{-i\lambda} 
\int_{-\infty}^\infty d\xi^1 \delta(\xi^1)= \frac{2\pi i}{\lambda}\ ,\ee
in agreement with \eq{fresnel}.

  Turning to the graded cyclic trace operation on 
$\Omega({\cal Z}_4)$, it is defined as 
\be 
{\rm STr}_{\Omega({\cal Z}_4)}\, f = \int_{{\cal R}^{\Real}_4} f \ ,
\label{TrZ}
\ee 
which projects onto the top form in $f$ and is compatible
with the hermitian conjugation operation and the outer automorphism
$\tau$, \emph{i.e.}
\be ({\rm STr}_{\Omega({\cal Z}_4)} f)^\dagger={\rm STr}\, f^\dagger\ ,\qquad
{\rm STr}_{\Omega({\cal Z}_4)} \tau(f)={\rm STr} f\ .\label{tracedf}\ee
Thus, if $f$ is a top form in $\Omega({\cal Z}_4)$ 
then its representative in $\Omega({\cal R}^{\Real}_4)$ 
must fall off sufficiently fast at infinity for the 
integral to be convergent. 
Thus ${\cal Z}_4$ must be a compact manifold obtained by 
adding points to ${\cal R}^{\Real}_4$ at infinity to extend 
its differential Poisson algebra structure
\cite{Chu:1997ik,Beggs:2003ne,McCurdy:2008ew,McCurdy:2009xz} 
(see also \cite{Arias:2015wha,Bonezzi:2015lfa}).
This can be achieved by assuming that ${\cal Z}_4$ admits 
a Poisson structure and a compatible pre-connection. 
We shall assume the latter to be trivial for simplicity.
In addition, we shall require ${\cal Z}_4$ to be closed, such that
\be {\rm STr}_{\Omega({\cal Z}_4)} \, df=0\ ,\label{TrZd}\ee
in order to avoid boundary terms from ${\cal Z}_4$ in varying the FCS action.
Assuming that $f$ and $g$ are two smooth symbols that fall off sufficiently
fast, it follows from \eq{starZI} that  
\be {\rm STr}_{\Omega({\cal Z}_{4})}\, f\star g=
\int_{{\cal R}^\Real_4} f\star g=\int_{{\cal R}^\Real_4} fg\ ,
\label{fstarg}\ee
which indeed implies the graded cyclicity, \emph{viz.}
\be {\rm STr}_{\Omega({\cal Z}_{4})}\, f\star g=
(-1)^{{\rm deg}(f){\rm deg}(g)}
{\rm STr}_{\Omega({\cal Z}_{4})}\, g\star f\ .
\label{cp}
\ee
As this property will be useful in analyzing boundary conditions
in the FCS model arising from the variational principle, we shall
assume that \eq{fstarg} holds for all elements in $\Omega({\cal Z}_{4})$
including distributions.
Finally, in order to obtain Vasiliev's equations from the FCS action, 
we shall assume that $\Omega({\cal Z}_4)$ admits global 
$SL(2;\Comp)$ symmetry and contains the (globally defined) 
closed two-forms
\be 
j_z=-\frac{i}4 dz^\a dz_\a \kappa_z\ ,\qquad \bar j_{\bar z} =(j_z)^\dagger\ ,
\label{jz}
\ee
where the inner Klein operator
\be 
\kappa_z=2\pi \delta^2(z^\a)\ .\label{kappayz}
\ee

  A choice of topology that satisfies all of the requirements as stated 
above in Eqs. \eq{TrZd}, \eq{fstarg}, \eq{cp}, \eq{tracedf}
and the inclusion the elements in \eq{jz}, is given by\footnote{The manifolds $S^4$ and $S^3\times S^1$ 
do not admit $j$, while $T^2 \times T^2$ breaks global 
$SL(2;\Comp)$ symmetry.}
\be \Omega({\cal Z}_4)= \bigoplus_{m,\bar m=0,1}
(\Omega(S^2)\star (j_z)^{\star m} )\otimes (\Omega(S^2)\star ({\bar j}_{\bar z})^{\star {\bar m}})\ ,\ee
where $\Omega(S^2)$ consists of globally defined 
forms on $S^2$ with Poisson structure obtained 
by extending the Poisson structure of \eq{starZ}
to the point at $\infty$.
At this point, the resulting Poisson bivector and 
all its derivatives vanish.
Hence, provided it is possible to exchange the
order of differentiation and summation in \eq{starZ} 
and using the fact that increasing number of derivatives 
of a form that falls off yields forms that fall off even 
faster, it follows that if $f,g\in \Omega(S^2)$ then
\be 
(f\star g)|_\infty=f|_\infty g|_\infty= 
(-1)^{{\rm deg}(f) {\rm deg}(g)} g|_\infty f|_\infty=
(-1)^{{\rm deg}(g) {\rm deg}(f)}(g\star f)|_\infty\ ,
\ee
\emph{i.e.} the point at infinity is a commuting 
point of $\Omega(S^2)$.
In other words, we are working with a topological
two-sphere equipped with a differential Poisson
algebra with trivial pre-connection.
Moreover, in order for the elements in $\Omega({\cal Z}_4)$
to have finite traces, we assume that the top forms on each
two-sphere fall off sufficiently fast at infinity working
in the original $\Real^2\times \Real^2$ coordinate chart.
For this fall-off condition to be embeddable in a differential
star product algebra, also the forms in lower degrees must
fall off appropriately at infinity. 
In particular, the only forms that can have finite values at infinity
are the zero-forms. 
Thus, in effect, we have taken  
\be 
{\cal Z}_4=S^2 \times S^2\ ,
\ee
by assuming boundary conditions at the commuting points 
at $\infty$ and allowing for delta function distributions 
at the origins so as to create a space of forms that is closed 
under exterior differentiation and star products, and has a
space of top forms with finite traces that vanish for
exact elements and obey \eq{fstarg}.
For this to hold true it is important that the delta function
$\kappa_z$ always appears together with line elements in the 
combination $j_z$ given in \eq{jz}, which obeys $j_z\star j_z=0$, 
whereas the inclusion of $\kappa_z$ into the algebra would 
require the inclusion of $j_z\star \kappa_z$ as well
which is not integrable.

In order to define the FCS model, we need to 
extend $\Omega({\cal Z}_4)$ into the algebra 
$\Omega({\cal Z}_4(k_z,\bar k_{\bar z}))$ by adding the
Klein operators $(k_z,\bar k_{\bar z})$ obeying
\begin{eqnarray}
& \{k_z, z_\a\}_\star = 0 = \{\bar k_{\bar z},\bar z_{\ad}\}_\star\ ,\qquad k_z\star k_z=
\bar k_{\bar z}\star \bar k_{\bar z}=1\ , &
\nonumber \\
&  [\bar k_{\bar z}, z_\a]_\star=0=[k_z,\bar z_{\ad}]_\star\ , 
\qquad \qquad  [k_z,\bar k_{\bar z}]_\star =0\ ,&
\nonumber \\
&  dk_z=k_zd\ ,\qquad d\bar k_{\bar z}=\bar k_{\bar z} d\ , &
\label{choice2}
\end{eqnarray}
Thus, the space
\be \Omega({\cal Z}_4(k_z,\bar k_{\bar z}))=\bigoplus_{m,\bar m=0,1} 
\Omega({\cal Z}_4)_{m,\bar m} \star (k_z)^{\star m}\star (\bar k_{\bar z})^{\star \bar m}\ ,\label{Omegamm}\ee
whose elements can thus be expanded as
\be f=\sum_{m,\bar m=0,1}
(k_z)^{\star m} \star (\bar k_{\bar z})^{\star \bar m}\star f_{m,\bar m}\ ,\qquad 
f_{m,\bar m}\in \Omega({\cal Z}_4)\ .\ee
It follows that
\be k_z\star f \star k_z= \pi_z( f)\ ,
\qquad \bar k_{\bar z}\star  f \star \bar k_{\bar z}
= \bar \pi_{\bar z}(f)\ ,\ee
where the reflection maps 
\be 
\pi_z: (z^\a,\bar z^{\ad})\rightarrow (-z^\a,\bar z^{\ad})\ ,\qquad
 \bar\pi_{\bar z}: (z^\a,\bar z^{\ad})\rightarrow (z^\a,-\bar z^{\ad})\ ,
 \label{piz}
 \ee
obey $\pi_z d=d\pi_z$ \emph{idem} for $\bar \pi_{\bar z}$ and act
as automorphisms of $\Omega({\cal Z}_4(k_z,\bar k_{\bar z}))$.
The graded cyclic trace operation can be extended from $\Omega({\cal Z}_4)$ to 
$\Omega({\cal Z}_4(k_z,\bar k_{\bar z}))$ by defining 
\be {\rm STr}_{\Omega({\cal Z}_4(k_z,\bar k_{\bar z}))} f= {\rm STr}_{\Omega({\cal Z}_4)} f_{0,0}\ .
\label{STrOmegaZ}\ee
To show the graded cyclicity we use that if $f\in \Omega({\cal Z}_4)$ then
\be 
{\rm STr}_{\Omega({\cal Z}_4)} \pi_z(f)={\rm STr}_{\Omega({\cal Z}_4)} \bar \pi_{\bar z}(f)=
{\rm STr}_{\Omega({\cal Z}_4)}f\ ,
\label{blaz}
\ee
from which it follows that if $f,g\in \Omega({\cal Z}_4(k_z,\bar k_{\bar z}))$ then
\bea 
{\rm STr}_{\Omega({\cal Z}_4(k_z,\bar k_{\bar z}))} f\star g
&=&
{\rm STr}_{\Omega({\cal Z}_4(k_z,\bar k_{\bar z}))}\sum_{m,\bar m} (k_z)^{\star m}\star
(\bar k_{\bar z})^{\star \bar m}\star f_{m,\bar m}
\star \sum_{n,\bar n=0,1} (k_z)^{\star n}\star
(\bar k_{\bar z})^{\star \bar n}\star g_{n,\bar n}\nn\\& =& \sum_{m,\bar n} {\rm STr}_{\Omega({\cal Z}_4)} 
(\pi_z)^m (\bar\pi_{\bar z})^{\bar m}(f_{m,\bar m})\star g_{m,\bar m}
\nn\\
&=& (-1)^{{\rm deg}(f){\rm deg}(g)}\sum_{m,\bar m} {\rm STr}_{\Omega({\cal Z}_4)}  g_{m,\bar m} \star (\pi_z)^{m} 
(\bar\pi_{\bar z})^{\bar m} (f_{m,\bar m})
\nn\\
&=&  (-1)^{{\rm deg}(f){\rm deg}(g)}\sum_{m,\bar m} {\rm STr}_{\Omega({\cal Z}_4)}  (\pi_z)^{m} (\bar\pi_{\bar z})^{\bar m}( g_{m,\bar m}) 
\star f_{m,\bar m}
\nn\\
&=& (-1)^{{\rm deg}(f){\rm deg}(g)}{\rm STr}_{\Omega({\cal Z}_4(k_z,\bar k_{\bar z}))} g\star f\ \label{cpz},
\eea
where the third and fourth equalities, respectively, follow from
\eq{cp} and \eq{blaz}. 
%


\section{Higher spin algebra}
\label{sec:HigherSpinAlgebra}

The FCS model will be formulated in terms of a single master
field, consisting of forms of different degrees,
valued in the direct product of a $\mathbb Z_2$-graded Frobenius 
algebra and an extension of the Weyl algebra by
inner Klein operators.
The FCS action also requires a trace operation
on these algebras.
%

\subsection{Matrix Frobenius algebra ${\cal F}$}

A Frobenius algebra ${\cal F}$ is a unital associative
algebra with non-degenerate bilinear form $\langle \cdot,\cdot\rangle_{\cal F}$
obeying the invariance property
$\langle M_1, M_2 M_3\rangle_{\cal F}
=\langle M_1 M_2, M_3\rangle_{\cal F}$.
In addition, we shall assume that $\langle\cdot,\cdot\rangle_{\cal F}$ is
realized as a trace operation, \emph{viz.}
$\langle M_1, M_2 \rangle_{\cal F}= {\rm Tr}_{\cal F} M_1 M_2$,
denoting the product in ${\cal F}$ by juxtaposition
(as we shall not need symbol calculus in ${\cal F}$).
Moreover, we shall assume that the Frobenius algebra contains
an inner Klein operator $h$ inducing a decomposition
into even and odd elements, \emph{viz.}
\be {\cal F}={\cal F}_+\oplus {\cal F}_-\ ,\qquad
h {\cal F}_\pm h=\pm {\cal F}_\pm\ ,\qquad
h^2=1\ .\ee
In order to decompose the FCS master field into 
coordinates and momenta, we shall use
the polarization induced by the decomposition
\be {\cal F}={\cal F}_0\oplus h {\cal F}_0\ ,\qquad
{\rm Tr}_{{\cal F}} \, h{\cal F}_0=0\ ,\label{f0}\ee
where ${\cal F}_0={\cal F}_{0,+}\oplus {\cal F}_{0,-}$
is thus a $\mathbb Z_2$-graded Frobenius subalgebra of ${\cal F}$
in which $h$ acts as an outer Klein operator, \emph{viz.}
\be h{\cal F}_{0,\pm}h=\pm{\cal F}_{0,\pm}\ .\ee
As will be explained elsewhere \cite{GenFCS},
the appearance of Frobenius algebras is a feature of a 
broader class of higher spin gauge theories based on 
generalized Chern--Simons actions, including the present model.

In our models, we shall use the eight-dimensional 
Frobenius algebra obtained by taking
\be {\cal F}_0={\rm Mat}_2(\Comp)=\oplus_{i,j=1,2} \Comp\otimes e_{ij}\ ,\qquad
e_{ij} e_{kl} = \delta_{jk}\,e_{il}\ ,\label{calF}\ee
and taking $h$ to be defined by
\be h^2=1\ ,\qquad h e_{ij} h= (e_{11}-e_{22}) e_{ij} (e_{11}-e_{22})\ ,
\label{falg}
\ee
or more explicitly,
\be [h , e_{11} ] = [h , e_{22} ]  =
\{ h , e_{12}\} = \{h,e_{21}\}=0\ ,\ee
that is, the adjoint action of $h$ is isomorphic to
that of $e_{11}-e_{22}$.
More compactly, we shall denote the basis elements of ${\cal F}$
by ($I,P=1,\dots,4$)
\be
e_I = (h e, h  \wt e, e,\wt{e} )   \ ,\qquad \quad f_P =  (h f , h  \wt f , 
f,\wt f) \ ,\qquad (e,\tilde e,f,\tilde f)=(e_{11},e_{22},e_{12},e_{21})\ .
\label{basisMat2h}
\ee
The algebra admits an additional three-grading
\be
{\cal F}={\cal F}^{(-1)}\oplus
{\cal F}^{(0)}\oplus {\cal F}^{(+1)}\ ,\qquad (e_{ij}, h\, e_{ij})\in {\cal F}^{(i-j)}\ .
\label{3grade}\ee
The hermitian conjugation and outer automorphism to be used are defined by
\be (e_{ij})^\dagger=e_{ji}\ ,\qquad h^\dagger =h\ ,\ee
\be \tau(e_{ij})=(f+\tilde f) e_{ij} (f+\tilde f)\ ,\qquad \tau(h)=-h\ ,\ee
or more explicitly,
\be (e,\tilde e,f,\tilde f;h)^\dagger=(e,\tilde e,\tilde f,f;h)\ ,\qquad
\tau(e,\tilde e,f,\tilde f;h)=(\tilde e,e,f,\tilde f;-h)\ .\ee
Finally, the trace operation on ${\cal F}$ compatible with these maps 
is taken to be
\be 
{\rm Tr}_{{\cal F}} \sum_{i,j} e_{ij} M^{ij}(h)=M^{11}(0)+
M^{22}(0)\ .
\label{tF}
\ee
%

\subsection{Extended Weyl algebra ${\cal W}$ }

The FCS model which will be presented in the next section requires 
the following extended Weyl algebra
\be 
{\cal W} =\bigoplus_{m,\bar m, r, \bar r =0,1} 
{\rm Aq}(2) \star (\kappa_y)^{\star r} \star (\bar \kappa_{\bar y})^{\star \bar r}
\label{calAmm}
\ee
where $Aq(2)$ consists of star polynomials in two complex doublets $(y^\a,\bar y^{\ad})$, $\a,\ad=1,2$, obeying the oscillator algebra
\be
[ y^\a, y^\b]_\star= 2i\e^{\a\b}\ ,\quad [ y^\a, \bar y^{\bd}]_\star=0\ ,
\quad
[ {\bar{y}}^{\ad},{\bar{y}}^{\bd}]_\star = 2 i\e^{\ad\bd}\ ,
\label{oscalg}
\ee
and the inner Klein operators are defined in Weyl order by the symbols 
\be 
\kappa_y:= 2\pi \delta^2(y)\ ,\qquad
\bar\kappa_{\bar y}:= 2\pi \delta^2(\bar y)\ .
\ee
These operators obey
\be 
\kappa_y\star P\star \kappa_y=\pi_y(P)\ ,\qquad
\bar\kappa_{\bar y}\star P\star \bar\kappa_{\bar y}=\bar\pi_{\bar y}(P)\ ,
\ee
where $\pi_y$ and $\bar\pi_{\bar y}$ are inner automorphisms  
whose action on symbols defined in Weyl order is given by
\be 
\pi_y (y) =-y\ ,\qquad  \bar\pi_{\yb} (\yb) = -\yb\ .
\label{py}
\ee
As an intermediate step, it is convenient to introduce 
also two outer Klein operators $(k_y,\bar k_{\bar y})$ 
obeying
\be 
\{ k_y, y^\a\}_\star =0= \{{\bar k}_{\bar y},{\bar{y}}^{\ad}\}_\star\ ,
\qquad k_y\star k_y={\bar k}_{\bar y}\star \bar k_{\bar y}=1\ ,
\label{Halg}\ee
\be [\bar k_{\bar y}, y^\a]_\star =0= [k_y,{\bar{y}}^{\ad}]_\star\ ,
\qquad [k_y,{\bar k}_{\bar y}]_\star =0\ ,
\ee
and defined the algebra
\be 
{\cal W}(k_y,\bar k_{\bar y}) =\bigoplus_{m,\bar m, r, \bar r =0,1} 
{\rm Aq}(2) \star (k_y)^{\star m}  \star ( \bar k_{\bar y})^{\star \bar m} 
\star (\kappa_y)^{\star r}
\star (\bar \kappa_{\bar y})^{\star \bar r}\ .
\ee
The generic elements of ${\cal W}(k_y,\bar k_{\bar y})$ are thus of the form
\be 
P=\sum_{r,\bar r=0,1} (\kappa_y)^{\star r}\star
(\bar\kappa_{\bar y})^{\star \bar r} \star P^{r,\bar r}(y,\bar y;k_y,\bar k_{\bar y})\ ,
\label{calA}
\ee
\be 
P^{r,\bar r}(y,\bar y;k_y,\bar k_{\bar y})=
\sum_{m,\bar m=0,1}  (k_y)^{\star m}\star 
(\bar k_{\bar y})^{\star\bar m} \star P^{r,\bar r}_{m,\bar m}(y,\bar y)\ ,
\nn
\ee
where $ P^{r,\bar r}_{m,\bar m}(y,\bar y)$ are Weyl ordered polynomials
\be
P_{m,\bar m}^{r,\bar r} (y,\bar y)
=\sum_{n,\bar n\geqslant 0} P^{r,\bar r}_{m,\bar m;\a(n),\ad(\bar n)}
y^{(\a_1}\star \cdots\star y^{\a_n)}\star 
\bar y^{(\ad_1}\star\cdots\star \bar y^{\ad_{\bar n})}\ .
\ee
%

\subsection{Trace operations}

Introducing the degree map ${\rm deg}_{\cal W}$ defined by 
\be 
k_y\star \bar k_{\bar y}\star P\star k_y\star \bar k_{\bar y}
=(-1)^{{\rm deg}_{\cal W}(P)} P\qquad\mbox{for $P\in {\cal W}(k_y,\bar k_{\bar y})$}\ .
\ee
one may define the supertrace operation
\be
{\rm STr}_{{\cal W}(k_y,\bar k_{\bar y})}\ P =P{}^{0,0}_{0,0}(0,0)\ ,
\label{step2}
\ee
whose restriction to ${\cal W}$ we denote by ${\rm STr}_{\cal W}$.
To show their graded cycliscity, we use the following lemmas 
\be 
{\rm STr}_{{\cal W}(k_y,\bar k_{\bar y})}\ P\star Q= (-1)^{{\rm deg}_{\cal W}(P){\rm deg}_{\cal W}(Q)}
{\rm STr}_{{\cal W}(k_y,\bar k_{\bar y})}\ Q\star P\qquad
\mbox{for\ \  $P,Q\in {\rm Aq}
(2) \star(k_y)^{\star m} \star (\bar k_\yb)^{\star n}$}\ .
\label{blay}
\ee
It follows that if $P,Q\in {\cal W}(k_y,\bar k_{\bar y})$ then
\bea
{\rm STr}_{{\cal W}(k_y,\bar k_{\bar y})}\ P\star Q&=& 
{\rm STr}_{{\cal W}(k_y,\bar k_{\bar y})}\ \left(\sum_{r,\bar r=0,1} (\kappa_y)^{\star r}\star
(\bar\kappa_{\bar y})^{\star \bar r}\star P^{r,\bar r}\right)
\star \left(\sum_{s,\bar s=0,1} (\kappa_y)^{\star s}\star
(\bar\kappa_{\bar y})^{\star \bar s}\star Q^{s,\bar s}\right)
\nn\\
& =& \sum_{r,\bar r=0,1} {\rm STr}_{\cal W}\ 
(\pi_y)^r (\bar\pi_{\bar y})^{\bar r}(P^{r,\bar r})\star Q^{r,\bar r}
\nn\\
&=&  (-1)^{{\rm deg}_{\cal W}(P){\rm deg}_{\cal W}(Q)}\sum_{r,\bar r=0,1} {\rm STr}_{\cal W}\ 
Q^{r,\bar r} \star (\pi_y)^{r}  (\bar\pi_{\bar y})^{\bar r} (P^{r,\bar r})
\nn\\
&=&  (-1)^{{\rm deg}_{\cal W}(P){\rm deg}_{\cal W}(Q)}\sum_{r,\bar r=0,1} {\rm STr}_{\cal W}\ 
(\pi_y)^{r} (\bar\pi_{\bar y})^{\bar r}( Q^{r,\bar r}) 
\star P^{r,\bar r}
\nn\\
&=& (-1)^{{\rm deg}_{\cal W}(P){\rm deg}_{\cal W}(Q)}{\rm STr}_{{\cal W}(k_y,\bar k_{\bar y})}\ Q\star P\ ,
\label{cpy}
\eea
where the third and fourth equalities, respectively, follow
from \eq{blay} and the fact that 
\be
{\rm STr}_{{\cal W}(k_y,\bar k_{\bar y})}\ \pi_y(P)
={\rm STr}_{{\cal W}(k_y,\bar k_{\bar y})}\ \bar\pi_{\bar y}(P)
= {\rm STr}_{{\cal W}(k_y,\bar k_{\bar y})}\ P\ . 
\ee
For the construction of the FCS model, it is important that 
the algebra ${\cal W}(k_y,\bar k_{\bar y})$ also admits 
the trace operation
\be 
{\rm Tr}_{{\cal W}(k_y,\bar k_{\bar y})}\ P :=  
{\rm STr}_{{\cal W}(k_y,\bar k_{\bar y})}\  \kappa_y \star \bar\kappa_{\bar y}\star P= P^{1,1}_{0,0}(0,0)\ ,
\label{TrA}
\ee
using \eq{calA} and \eq{step2}, which is thus equivalent to the
projection of $P$ onto its component along the generator 
$\kappa_y \star \bar\kappa_{\bar y}$.
The trace operation is nondegenerate, \emph{i.e.} $P$ vanishes if 
${\rm Tr}_{{\cal W}(k_y,\bar k_{\bar y})}\ P\star Q=0$ for all $Q$, and enjoys the following properties:
\be 
({\rm Tr}_{{\cal W}(k_y,\bar k_{\bar y})}\ P)^\dagger= 
{\rm Tr}_{{\cal W}(k_y,\bar k_{\bar y})}\ P^\dagger\ ,
\ee
\be 
{\rm Tr}_{{\cal W}(k_y,\bar k_{\bar y})}\ \tau(P)=
{\rm Tr}_{{\cal W}(k_y,\bar k_{\bar y})}\ \pi_y(P)
={\rm Tr}_{{\cal W}(k_y,\bar k_{\bar y})}\ \bar\pi_{\bar y}(P)= 
{\rm Tr}_{{\cal W}(k_y,\bar k_{\bar y})}\ P\ .
\label{compTrA}
\ee
where the hermitian conjugation is defined as
\be 
(y^\a,\bar y^{\bd}; k_y,\bar k_{\bar y})^\dagger 
=(\bar y^{\ad},y^\b;\bar k_{\bar y}, k_y)\ .
\ee
and the outer $\tau $--automorphism as
\be 
\tau(y^\a,\bar y^{\ad}; k_y,\bar k_{\bar y})
=(iy^\a,i\bar y^{\ad}; k_y,\bar k_{\bar y})\ .
\ee
%

\section{Frobenius--Chern--Simons model}
\label{sec:FCSmodel}

In this section, we use the ingredients introduced so far to
construct a covariant Hamiltonian action in nine dimensions 
consisting of a Chern--Simons-like bulk action and a boundary 
term, whose definition requires the choice of a polarization.
In Section 4.1 we consider field configurations 
that are defined globally on generalized bundle
spaces.
Locally defined configurations are considered in Section 4.2.
We then turn to a convenient component formulation
and kinematic constraints.


\subsection{Superconnection and the action}


In this section we consider globally defined configurations given 
by sections in 
\footnote{Various consistent truncations of the system will be discussed in Section 4.4. 
The definition of trace operations below will straightforwardly carry over to those truncated systems.} 
\be 
{\cal E}= \tfrac{1}{16} (1+\pi\bar\pi) 
(1+\pi_{k_y}\pi_{k_z})(1+\bar\pi_{\bar k_{\bar y}}\bar\pi_{\bar k_{\bar z}})
\Big(\left[\Omega({\cal X}_5)\otimes \Omega({\cal Z}_4(k_z,\bar k_{\bar z})) \otimes {\cal F}
\otimes{\cal W}(k_y,\bar k_{\bar y}) \right]\star (1+k\star \bar k)\Big)\ ,
\label{varE}
\ee
where $\pi_{k_y}$ is the outer automorphisms that maps $k_y$ to $-k_y$,
\emph{idem} $\pi_{k_z}$, $\bar\pi_{\bar k_{\bar y}}$ and $\bar\pi_{\bar k_{\bar z}}$, and 
\be 
k = k_y \k_z\ ,\qquad \bar k =\bar k_{\bar y} \bar k_{\bar z}\ ,\qquad \pi=\pi_y\pi_z\ ,\qquad \bar\pi=\bar\pi_{\bar y} \bar\pi_{\bar z}\ ,
\label{pibarpi}
\ee
where $(\pi_z, \bar\pi_{\bar z})$ are defined in \eq{piz} and 
$(\pi_y, \bar\pi_{\bar y})$ are from \eq{py}.
The following relations hold
\be k\star f\star k= \pi(f)\ ,\qquad \bar k\star f\star \bar k= \bar \pi(f)\ .\ee
The projection by $\frac{1}{4} (1+\pi_{k_y}\pi_{k_z})(1+\bar\pi_{\bar k_{\bar y}}\bar\pi_{\bar k_{\bar z}})$ ensures that the dependence 
on the outer Klein operators is only via $k$ and $\bar k$,
which is natural from the point of view of a possible connection with topological open string theory. 
The projection by $\frac{1}{2} (1+\pi\bar\pi)$ keeps only integer spin fields,
as we considering bosonic models.
Finally, the $\star$-product projection by $\frac12 (1+k\star \bar k)$ is employed 
in order to avoid a duplication into two copies of the theory. 
We may simplify \eq{varE} as follows:
\be 
{\cal E}=\Omega({\cal X}_5)\otimes \tfrac14 (1+\pi \bar\pi) \Big(\left[ \Omega({\cal Z}_4) \otimes  {\cal A}\right]\star (1+k\star\bar k)\Big) \ ,
\ee
where we have defined
\be 
{\cal A} = {\cal F} \otimes{\cal W}\otimes {\cal K} \ , 
\qquad {\cal K}=\{ 1, k, {\bar k} , k \star {\bar k} \}\ ,
\label{defcalA}
\ee
and ${\cal W}$ is the extended Weyl defined in \eq{calAmm}.
The sections in ${\cal E}$ can be seen as sections of 
a bundle over ${\cal X}_5$ with fiber given by $\frac14 (1+\pi \bar\pi) \Big(\left[ \Omega({\cal Z}_4) \otimes  {\cal A}\right]\star (1+k\star\bar k)\Big)$,
which is a graded associative differential algebra with degree
and differential given by the form degree and exterior derivative
on ${\cal Z}_4$, respectively.
Alternatively, they can be seen as $\frac14 (1+\pi \bar\pi)$-projected 
sections of a generalized bundle over ${\cal X}_5\otimes {\cal Z}_4$
with fiber given by $\tfrac12 {\cal A}\star (1+k\star\bar k)$.
More precisely, introducing a basis $T^\Lambda$ of ${\cal A}$,
an element $f\in {\cal E}$ can be expanded as
\be 
f=\sum_\Lambda f_\Lambda \star T^\Lambda\ ,\qquad f_\Lambda\in \Omega({\cal X}_5)\otimes \Omega({\cal Z}_4) \ ,\label{expandf}
\ee
where the dependences of $f_\Lambda$ on $(z^\a,\bar z^{\ad})$ and $T^\Lambda$ on
$(y^\a,\bar y^{\ad})$ are in terms of symbols defined using Weyl order.
Assuming furthermore that $T^\Lambda=T^\Lambda_{m_\Lambda,\bar m_\Lambda}
\star k^{m_\Lambda}\star \bar k^{\bar m_\Lambda}$, it follows that 
$f_\Lambda\star T^{\Lambda'}= T^{\Lambda'}\star (\pi_z)^{m_\Lambda}
(\bar \pi_{\bar z})^{\bar m_\Lambda} f_\Lambda$, that is, the
fiber and base elements are not entirely separated from each other.

For $f\in \cal E$, we define
\be   {\rm Tr}_{{\cal E}} f  := \int_{{\cal M}_9} {\rm Tr}_{\cal A}\ f\ , \qquad 
{\rm Tr}_{\cal A}:= {\rm Tr}_{\cal F}\, {\rm Tr}_{{\cal W}}\, {\rm Tr}_{\cal K}\ ,
\ee
where ${\rm Tr}_{\cal F}$ is defined in \eq{tF}, ${\rm Tr}_{{\cal W}}$ in
 \eq{TrA} (upon setting $k_y=\bar k_{\bar y}=0$), and 
\be {\rm Tr}_{\cal K} f:=f|_{k=0=\bar k}\ .\ee
From the odd dimensionality of the base manifold, and 
the relations 
\be 
{\rm Tr}_{{\cal E}}\ \pi(f)={\rm Tr}_{{\cal E}}\ \bar\pi(f)={\rm Tr}_{{\cal E}}\ f\ ,
\label{blayz}
\ee
that hold for any $f\in {\cal E}$, as can be seen by combining \eq{blaz} and \eq{blay}, it follows that ${\rm Tr}_{{\cal E}}$ is cyclic: 
\bea 
{\rm Tr}_{{\cal E}}\ f\star g
&=&
{\rm Tr}_{{\cal E}}\ \sum_{m,\bar m} k^{\star m}\star
\bar k^{\star \bar m}\star f_{m,\bar m}
\star \sum_{n,\bar n=0,1} k^{\star n}\star
\bar k^{\star \bar n}\star g_{n,\bar n}
\nn\\
& =& \sum_{m,\bar n} {\rm Tr}_{{\cal E}}\
\pi^m \bar\pi^{\bar m}(f_{m,\bar m})\star g_{m,\bar m}
\nn\\
&=& \sum_{m,\bar m} {\rm Tr}_{{\cal E}}\  g_{m,\bar m} \star \pi^{m} 
\bar\pi^{\bar m} (f_{m,\bar m})
\nn\\
&=&  \sum_{m,\bar m}{\rm Tr}_{{\cal E}}\
\pi^{m} \bar\pi^{\bar m}( g_{m,\bar m}) 
\star f_{m,\bar m}={\rm Tr}_{{\cal E}}\  g\star f\ ,
\eea
where the third equality requires \eq{cpz}, \eq{TrA} and the fact
that if $f$ is an even form on ${\cal M}_9$ then $g$ is even and vice 
versa, while the fourth equality requires \eq{blayz}. 
In terms of the expansion \eq{expandf}, the trace operation reads
\be 
\int_{{\cal M}_9 } {\rm Tr}_{\cal A}\  f=\sum_{\Lambda} \int_{{\cal M}_9 }f_\Lambda   
\left({\rm Tr}_{\cal A} T^\Lambda\right)=
\sum_{\Lambda} \int_{{\cal X}_5} \int_{{\cal Z}_4 } f_\Lambda 
\left({\rm Tr}_{\cal A}T^\Lambda\right)\,
\ .
\label{basis2}
\ee
Thus, prior to summing over $\Lambda$ and integrating over ${\cal X}_5$, 
the combined $\int_{{\cal Z}_4}{\rm Tr}_{\cal A}$ operation yields a finite 
result.

Introducing a superdegree ${\rm deg}_{\cal E}$ on ${\cal E}$ given 
by the sum of the form degree on ${\cal M}_9$ and the 3-degree in 
${\cal F}$, and letting ${\cal E}_+$ and ${\cal E}_-$ denote the 
projections of ${\cal E}$ onto its subspaces with even and
odd superdegree, respectively, that is
\be 
{\cal E}_\pm = \bigoplus_{\sigma=\pm}\tfrac14 (1+\pi \bar\pi) 
\Big(\left[\Omega({\cal M}_9)_{\sigma} \otimes {\cal F}_{\pm\sigma}
\otimes{{\cal W}}\otimes {\cal K}\right]\star (1+k\star\bar k)\Big) \ ,
\ee
where $\Omega({\cal M}_9)_+$ and $\Omega({\cal M}_9)_-$ 
denote the spaces of even and odd forms on ${\cal M}_9$, 
respectively,
it follows that ${\cal E}$ acquires the structure of a 
graded associative differential algebra with superdifferential
\be q:=hd\ \ee
Indeed, $q:{\cal E}_\pm \rightarrow {\cal E}_\mp$ and 
\be 
q(f\star g)=qf\star g+(-1)^{{\rm deg}_{\cal E}\, f} f\star qg\ ,\qquad {\rm Tr}_{{\cal E}}\ qf=
\oint_{\partial{\cal M}_9} {\rm Tr}_{\cal A}\ f\ .
\ee
Hence, introducing a globally defined superconnection 
\be 
Z\in {\cal E}_-\ .\ee
we can define the following action functional 
\be 
S = \int_{{\cal M}_9}  {\rm Tr}_{\cal A}
\left(\tfrac12 \, Z\star q Z+\tfrac13 \,Z\star Z\star Z\right)
-\frac{1}{4}\, \oint_{\partial{\cal M}_9} {\rm Tr}_{{\cal A}} \,[h\pi_h(Z) \star Z]
\label{FCS}
\ee
that we shall refer to as the Frobenius--Chern--Simons action.
The action gauge invariant and has a finite Lagrangian density 
on ${\cal M}_9$, given by the trace over ${\cal A}$.
The total variation 
\be \delta S = \int_{{\cal M}_9} 
{\rm Tr}_{\cal A}\ 
\delta Z\star R +\frac12 \oint_{{\cal M}_9} 
{\rm Tr}_{\cal A}\ h\,\delta Z\star Z\ ,
\label{deltaFCS}\ee
where the Cartan curvature
\be
R := qZ + Z\star Z \in {\cal E}_+\ . \label{R}
\ee
The equation of motion $R=0$ is Cartan integrable, hence 
transforming covariantly under the gauge transformations
\be
 \delta  Z =  {q}   \theta +[ Z, \theta]_\star\ ,\qquad
\delta  R=[ R, \theta]_\star\ .
\ee
Perturbatively, $Z$ can be given on-shell in terms of a gauge function 
$L$ (containing forms in different degrees) and an integration constant 
$C$, \emph{viz.}
\be Z= L^{\star(-1)}\star (q+C)\star L\ ,\qquad qC=C\star C=0\ ,\label{C&L}\ee
where the algebraic condition on $C$ is a consequence of \eq{topform}
which implies that $C\in{\cal F}^{(-1)}\otimes {\cal W}\otimes {\cal K}$.
More generally, as the base manifold is noncommutative, there exist
additional solutions that make use of projectors, which cannot
be expressed using gauge functions. 

The models defined above contain top forms on 
${\cal M}_9$, and, consequently, the equations
of motion contain (quadratic) zero-form constraints\footnote{
The treatment of FCS models with top forms,
for which the AKSZ formalism provides a natural framework, 
will be given elsewhere \cite{GenFCS}.}.
To avoid these constraints, we shall henceforth assume that
$Z$ does not contain any top forms and that its $8$-form and $0$-form
sectors do not contain any components in ${\cal F}^{(-1)}$ and
${\cal F}^{(+1)}$, respectively; put in equations, 
\be Z\cap \Omega_{[9]}({\cal M}_9)=0\ ,\qquad Z\cap 
\left(\Omega_{[8]}({\cal M}_9)\otimes {\cal F}^{(-1)}\right)=0\ ,
\qquad Z\cap \left(\Omega_{[0]}({\cal M}_9)\otimes {\cal F}^{(+1)}\right)=0\ ,\label{topform}\ee
using \eq{3grade}.
As a result, we have
\be R\cap \Omega_{[0]}({\cal M}_9)=0\ ,\ee
that is, the equations of motion do not contain any zero-form constraints.


\subsection{Global formulation }


%
Covering ${\cal X}_5$ by a set of charts ${\cal X}_{5,\xi}$, and 
letting ${\cal E}_{\xi,\pm}$ denote the restriction of ${\cal E}_\pm$
to locally defined sections over 
\be {\cal M}_{9,\xi}={\cal X}_{5,\xi}\times {\cal Z}_4\ ,\ee
we may construct an action functional for a set of local
representatives $Z_\xi\in {\cal E}_{\xi,-}$ by modifying
the action \eq{FCS} by (locally defined) total derivatives.
In view of the fact that 
\be \Omega\left({\cal M}_{9,\xi}\cap {\cal M}_{9,\eta}\right)=
\Omega\left({\cal X}_{5,\xi}\cap {\cal X}_{5,\eta}\right)\otimes \Omega({\cal Z}_4)\ ,\ee
one way to achieve this is to use \eq{f0} to define
\be {\cal E}={\cal E}_{\xi,0}\oplus h{\cal E}_{\xi,1}
\ ,\ee
where thus ${\cal E}_{\xi,0}$ and ${\cal E}_{\xi,1}$ 
are $h$-independent, and split 
\be
 Z_\xi = P_\xi+h X_\xi \ ,\qquad X_\xi,\,P_\xi \in
{\cal E}_{\xi,0,-}\ .\label{XP}
\ee
It follows that
\be
R_\xi = R^X_\xi + R^P_\xi\ ,\qquad R^X_\xi,\,R^P_\xi\in {\cal E}_{\xi,0,+}\ ,
\ee
where 
\be
R^X:=  F^X+P\star P\ ,\qquad R^P:= Q P\ ,
\label{RX}
\ee
and we have defined
\be 
F^X:=dX+hXh\star X\ ,\qquad Q P:=q P + hX\star P +P\star hX\ .
\ee
The differentials $q$ and $Q$ obey the graded Leibniz rule
\be
q(f\star g) = q(f)\star g + (-1)^{{\rm deg}_{\cal E} f} f\star q g\ ,
\qquad Q(f\star g) = Q(f)\star g + (-1)^{{\rm deg}_{\cal E} f} f\star Q g\ 
\ee
Likewise, splitting
\be
\theta_\xi = \e^X_\xi+ h\, \e^P_\xi\ ,\qquad \e^X_\xi,\,\e^P_\xi\in {\cal E}_{\xi,0,+}\ ,
\label{theta}
\ee
the gauge transformations take the form
\bea
\delta X &=& {d} \e^X + X\star \e^X - h {\e}^X h  \star X 
+ hPh\star \epsilon^P -\epsilon^P \star P\ ,
\nonumber \\
\delta P &=& {d} \e^P + h X h\star  \e^P - h \e^P h \star X+ [P,\e^X]_\star  \ .
\label{xpt}
\eea
Since $F^X$ and $QP$ transform homogeneously under gauge the
transformations with $\e^X$ parameters, we may construct
a globally defined Lagrangian by choosing a structure 
group with transition functions 
\be 
T^\xi_\eta \in {\cal E}_{\xi,\eta,0,+}\ ,
\ee
consisting of $h$-independent forms with even superdegree 
defined on the overlaps ${\cal M}_{9,\xi}\cap 
{\cal M}_{9,\eta}$, or a subgroup thereof.
It follows that 
\be X_\xi=hT^\eta_\xi h\star (X_\eta+{d})\star T^\xi_\eta\ ,
\qquad P_\xi=T^\eta_\xi\star P_\eta\star T^\xi_\eta\ .\label{gluingXP}\ee
and hence, ,
it follows that
\be F^X_\xi=hT^\eta_\xi h\star F^X_\eta\star T^\xi_\eta\ .\ee
An action functional that is well defined for locally
defined configurations and that reduces to the FCS action 
\eq{FCS} for globally defined configurations is given by
\be S = \sum_{\xi} \int_{{\cal M}'_{9,\xi}} 
{\rm Tr}_{\cal A}\
\Big[ \tfrac12 \, Z_\xi\star q Z_\xi
+\tfrac{1}{3}\, Z_\xi\star Z_\xi\star Z_\xi 
+\tfrac12 {d}\left( X_\xi\star P_\xi \right) \Big]\ ,
\ee
using patches\footnote{Alternatively, we could define 
$\int_{{\cal M}'_{9,\xi}}=\int_{{\cal M}_{9,\xi}}\rho_\xi$
using a partition $\rho_\xi$ of unity.} ${\cal M}'_{9,\xi}\subset {\cal M}_{9,\xi}$ 
such that $\bigcup_\xi {\cal M}'_{9,\xi}={\cal M}_9$.
It follows that  
\be S =\sum_\xi \int_{{\cal X}'_{5,\xi}} \check{\cal L}_\xi\ ,\ee
where
\be \check{\cal L}_\xi=\int_{{\cal Z}_4}{\rm Tr}_{\cal A}\
\left(P_\xi \star F^X_\xi +\tfrac13 P_\xi\star P_\xi\star P_\xi\right)\in \Omega({\cal X}_{5,\xi})\ .
\ee
In an overlap region one has
\bea 
\check{\cal L}_\xi&=& \int_{{\cal Z}_4}{\rm Tr}_{\cal A}\
T^\eta_\xi\star \left(P_\eta \star F^X_\eta +\tfrac13 P_\eta\star P_\eta\star P_\eta\right)\star T^\xi_\eta
\nn\w2
&=&\int_{{\cal Z}_4}{\rm Tr}_{\cal A}\
\left(P_\eta \star F^X_\eta +\tfrac13 P_\eta\star P_\eta\star P_\eta\right)
=\check{\cal L}_\eta\ ,
\eea
that is, there exists a globally defined form 
$\check{\cal L} \in \Omega({\cal X}_5)$ such that
\be \check{\cal L}_\xi=\check{\cal L}|_\xi\ ,\ee
which is to say that $S$ is globally defined.
For simplicity, we write
\be 
S  =\int_{{\cal M}_9} {\rm Tr}_{\cal A}\ \left(P \star F^X
+\tfrac13 P\star P\star P\right)\ ,
\label{SBulk}
\ee
where thus $F^X=dX+hXh\star X$ and the combined 
$\int_{{\cal Z}_4} {\rm Tr}_{\cal A}$ 
operation yields a globally defined top form on ${\cal X}_5$. 
This action is invariant under the gauge transformations \eq{xpt}.

Two modifications remain to be performed:
First, we will choose\footnote{Whether there exist nontrivial
transition functions on ${\cal Z}_4$ remains to be investigated.}
\be 
T^\xi_\eta \in \tfrac14(1+\pi\bar\pi) \Big(\left[\Omega\left({\cal X}_{5,\xi}
\cap {\cal X}_{5,\eta}\right)\otimes
1_{\Omega({\cal Z}_{4})} \otimes {\cal F}_0\otimes 
{{\cal W}}\otimes {\cal K}\right]\star (1+k\star \bar k)\Big)_+\ ,
\label{Transition}
\ee
\emph{i.e.} a set of transition functions 
that are constant on ${\cal Z}_{4}$, then 
\be 
{\cal L}={\rm Tr}_{\cal A} \, \left(P \star F^X
+\tfrac13 P\star P\star P\right) \in \Omega({\cal X}_5)\otimes \Omega({\cal Z}_4)\ .
\ee
is a globally defined form on ${\cal M}_9$ such that
\be 
S  =\sum_\xi \int_{{\cal M}'_{9,\xi}} 
{\cal L}_\xi\ , \qquad {\cal L}_\xi={\cal L}|_\xi\ .
\ee
Second, as $X$ enters the action only via its curvature,
we can take 
\be 
X_\xi\in \tfrac14(1+\pi\bar\pi) \Big(\left[
\Omega({\cal X}_{5,\xi}) \otimes \Omega({\cal R}^\Real_4) 
\otimes {\cal F}_0\otimes {\cal W}\otimes {\cal K}\right]\star (1+k\star \bar k)\Big)_- \ ,
\label{Xxi}
\ee
that is, we allow $X$ to develop singularities at the
point at infinity of ${\cal Z}_4$ provided that
\be 
F^X_\xi\in {\cal E}_{\xi,0,+}\ .
\ee

The general variation, under the assumptions made above, takes the form
\be
\delta S  =  \int_{{\cal M}_9} 
{\rm Tr}_{\cal A}\  \left( \delta X \star R^P h
+ \delta P \star R^X +{d}(\delta X\star P)\right)\ ,
\label{total}
\ee
where the total derivatives cancel between neighboring patches (in the interior of ${\cal M}_9$)
since $\delta X$ and $P$ belong to sections due to \eq{gluingXP}, leaving 
\be 
\int_{{\cal M}_9} {\rm Tr}_{\cal A}\ 
d(\delta X\star P) = \oint_{\partial{\cal M}_9}{\rm Tr}_{\cal A}\ \delta X\star P\ .
\ee
As we are working under the assumption that the elements in $\Omega({\cal Z}_4)$
and ${\cal W}$ are given by symbols defined using the Weyl order, the above
quantity can be rewritten by using \eq{basis2} followed by \eq{fstarg} to 
replace the star product in $\Omega({\cal Z}_4)$ by the classical product 
(keeping in mind that $j_z\star j_z=0$),
\emph{viz.}
\be 
\int_{{\cal M}_9} {\rm Tr}_{\cal A}\ 
d(\delta X\star P) =\oint_{\partial{\cal M}_9}{\rm Tr}_{\cal A}\ \delta X
\star_{\cal A}\ P\ .
\ee
Hence, if $X$ is free to fluctuate at $\partial{\cal M}_9$,
it follows from the variational principle that
\be 
P|_{\partial {\cal M}_9}= 0\ .
\ee
Finally, whereas $\check{\cal L}$ is invariant (pointwise on ${\cal X}_5$) 
under gauge transformations with parameters $\epsilon^X$,
it transforms into a total derivative under transformations with
parameters $\epsilon^P$, \emph{viz.}
\be 
\delta_{\e^P} S =\int_{{\cal M}_9} {\rm Tr}_{\cal A}\ 
{d}\left(\e^P\star R^X\right)\ ,
\ee
that vanishes provided that $\e^P$ belongs to the same section as $P$ and
\be
\e^P|_{\partial{\cal M}_9}=0\ .
\ee


\subsection{Component formulation}
\label{Sec:4.3}

We expand $Z$ in the basis \eq{basisMat2h} for ${\cal F}$, using the notation
\be
Z= \sum_I  A^I e_I +\sum_P  B^P f_P\ ,\qquad
A^I \equiv ( A, {\wt A},  V,  {\wt V})\ ,\qquad
B^P \equiv (B, {\wt B},   U,   {\wt U})\ .
\label{comp1}
\ee
Recalling \eq{topform}, we have the form degrees 
\bea
& {\rm deg} (B,A,\widetilde A,\widetilde B)\in\left\{(2n,1+2n,1+2n,2+2n)\right\}_{n=0,1,2,3}\ ,&
\nonumber \\
& {\rm deg} (\widetilde U,V,\widetilde V,U)=\left\{(8-2n,7-2n,7-2n,6-2n)
\right\}_{n=0,1,2,3}\ ,&
\label{content2}
\eea
Thus, employing the basis \eq{basisMat2h} in \eq{comp1}, we get
\be
Z= hX + P\ ,
\ee
where
\be 
X = A e +  \wt A {\tilde e} + B f + \wt B {\tilde f}= 
\begin{pmatrix} A & B \\ \widetilde B & \widetilde A  \end{pmatrix}\ ,
\qquad 
P = V e + \wt V {\tilde e}+  U f +  \wt U {\tilde f} =
\begin{pmatrix} V & U \\  \wt U & \wt V \end{pmatrix}
\label{XYD}
\ee
The action \eq{FCS}, which can also be written as in \eq{SBulk}, takes the form
\bea
S &=&  \int_{{\cal M}_9} {\rm Tr}_{{\cal W}\otimes {\cal K}}\Big[
\wt U \star DB  + V\star \left( F -  B \star\wt B 
+ \tfrac{1}{3}\, V^{\star2}+ U \star\wt U \right)
\nn\\
&&  \ \ \ \ \ \ \ +U \star \wt D \wt B
+ \wt V\star \left( \wt F - \wt B\star B + \tfrac{1}{3}\, {\wt V}^{\star2} +  
\wt U\star U \right) \Big]\ ,
\label{SH}
\eea
making use of the definitions
\bea
& F :={d} A+A\star A\ ,\qquad\qquad\qquad\quad \widetilde F:=
{d} \widetilde A+\widetilde A\star \widetilde A\ ,&
\nonumber\\ 
& DB :={d} B+A\star B-B\star \widetilde A\ ,\qquad 
\widetilde D \widetilde B :={d} \widetilde B
+\widetilde A\star \widetilde B-\widetilde B\star A\ ,&
\nonumber\\ 
& D U :={d} U+A\star U-U\star \widetilde A\ ,\qquad 
\widetilde D \widetilde U :={d} \widetilde U+
\widetilde A\star \widetilde U-\widetilde U\star A\ ,&
\nonumber \\ 
& D V :={d} V+A\star V+V\star  A\ ,\qquad 
\widetilde D \widetilde V :={d}\widetilde V+
\widetilde A\star \widetilde V+\widetilde V\star \widetilde A\ .&
\eea
The bulk equations of motion, which amount to vanishing Cartan curvatures, read
\bea
& F-B\star\widetilde B  + V\star V+U\star \widetilde U=0
\ ,\qquad DB +V\star U + U\star \widetilde V=0\ ,&
\nonumber\\
& \widetilde F - \widetilde B\star B +
\widetilde V \star \widetilde V+\widetilde U \star U=0\ ,
\qquad  
\widetilde D \widetilde B + \widetilde V \star \widetilde U +\widetilde U\star V=0\ ,&
\nonumber\\
& D U +B\star \widetilde V+ V\star B=0 \ ,\qquad DV+B\star \widetilde U-U\star \widetilde B=0\ ,&
\nonumber\\
& \widetilde D\widetilde U +\widetilde B\star V+\widetilde V\star \widetilde B=0\ ,\qquad  
\widetilde D \widetilde V+\widetilde B\star U-\widetilde U\star B=0\ .&
\label{cc2}
\eea
In accordance with \eq{theta}, the gauge parameter can be written 
as $\theta= \e^X + h\e^P$, where we introduce the notation
\be
\e^X= \begin{pmatrix}   \e & \eta \\ \wt\eta & \wt\e \end{pmatrix}\ ,\qquad 
\e^P= \begin{pmatrix}   \e^V & \eta^U \\ \eta^{\wt U} &\e^{\wt V} \end{pmatrix}\ .
\label{exp}
\ee
For simplicity in notation, we have suppresses the superscripts $A$ and $B$ 
in the components of $\e^X$. The transformation rules for the component fields 
can be readily obtained from \eq{xpt} by substitution of \eq{XYD}.

Thus, on $\partial{\cal M}_9$, where $P$ and hence $(U,\widetilde U;V,\widetilde V)$
vanish, we arrive at\footnote{The Cartan integrability of eq. \eq{E2}
is formally equivalent to the Cartan integrability of the
equations of motion of the three-dimensional matter-coupled
higher spin gravities proposed in \cite{Fujisawa:2013ima}.
}
\bea
& F - B\star\widetilde B= 0\ ,\qquad D B= 0\ ,&
\nonumber\\
& \widetilde F - \widetilde B\star B= 0\ ,\qquad \widetilde D \widetilde B= 0\ ,&
\label{E2}
\eea
which is the desired modification of Vasiliev's original
system \cite{Vasiliev:1990en,Vasiliev:1992av}. 
By going to the basis,
\be
\wt A = W+K\ ,  \qquad A= W-K\ ,
\label{defW}
\ee
the equations of motion on the boundary can be written as
\bea
& F_W + K\star K - \frac12\{B,\widetilde B\}_\star= 0\ ,\qquad
D_W  K - \frac12[\widetilde B,B]_\star=0\ ,&
\nonumber\\
& D_W  B-\{K,B\}_\star=0\ ,\qquad
D_W \widetilde B+\{K,\widetilde B\}_\star= 0\ ,&
\label{eom2}
\eea
where we have defined $D_W f={d}f + W \star f-(-1)^{\rm deg(f)}f\star W$ 
and $F_W={d}W+W^2$.
Since $\e^P\vert_{\partial{\cal M}_9} = 0$, recalling the notation \eq{exp}, and 
splitting the gauge parameters $(\e,\wt\e)$ as
\be  \e = \alpha -\beta\ ,\qquad \wt\e = \alpha +\beta\ , \ee
the gauge transformations under which the field equations \eq{eom2} are invariant can 
be written as 
\bea
\delta W &=&  D_W\alpha +[K,\beta]_\star +\tfrac{1}{2}\, \{\wt\eta,B\}_\star +\tfrac{1}{2}\,\{\wt B,\eta\}_\star\ ,
\nonumber\\
\delta K &=& D_W\beta +[K,\alpha]_\star +\tfrac{1}{2}\,[\wt\eta,B]_\star + \tfrac{1}{2}\, [\wt B,\eta]_\star\ ,
\nonumber\\
\delta B &=& D_W\eta +[B,\alpha]_\star-[K,\eta]_\star  +\{B,\beta\}_\star\ ,
\nonumber\\
\delta \wt B &=& D_W\wt \eta +[\wt B,\alpha]_\star+[K,\wt\eta]_\star  -\{\wt B,\beta\}_\star\ .
\label{dt4}
\eea
%

\subsection{Minimal and augmented bosonic models}\label{Sec:2.8}

The FCS model can be truncated consistently off-shell 
by applying the following operations:
\\ 
i) The hermitian conjugation defined by
\be (y_\a,\yb_{\ad};z_\a,\zb_{\ad};k,\bar k ;e_{ij},h)^\dagger=(\yb_{\ad},y_\a;-\zb_{\ad},-z_\a;\bar k,k; e_{ji},h)\ ,\ee
\be (f\star g)^\dagger=(-1)^{{\rm deg}(f){\rm deg}(g)}
g^\dagger \star f^\dagger\ ,\qquad ({d}f)^\dagger= {d}(f^\dagger)\ ;\ee
ii) The linear anti-involution $\tau$ defined by 
 \be
 \tau(y_\a , \yb_{\ad} ; z_\a,\zb_{\ad};k,\bar k;e,\tilde e,f,\tilde f;h)=(i y_\a,i\yb_{\ad};-iz_\a,-i\zb_{\ad};k, \bar k;\tilde e,e,f,\tilde f;-h)\ ,\ee
\be \tau(f\star g)=(-1)^{{\rm deg}(f){\rm deg}(g)} \tau(g)\star\tau(f)\ ,\qquad
\tau({d}f)={d}(\tau(f))\ ;\ee
iii) The linear automorphisms $\pi_{k}$ and $\pi_{{\bar k}}$ defined by
\be \pi_{k}(k,\bar k)=(-k,\bar k)\ ,\qquad \pi_{k}(f\star g)=\pi_{k}(f)\star \pi_{k}(g)\ ,\qquad \pi_{k}({d}f)
={d}(\pi_{k}(f))\ ,
\label{pis}
\ee
while all other elements are unaffected, \emph{idem} $\pi_{{\bar k}}$.

Composing these maps with the combined operation of integrating
over ${\cal M}$ and tracing over ${\cal A}$ one 
has 
\be 
\left(\int_{{\cal M}}{\rm Tr}_{\cal A}\ f\right)^\dagger
=\int_{{\cal M}}{\rm Tr}_{\cal A}\ f^\dagger\ ,\ee\be
\int_{{\cal M}}{\rm Tr}_{\cal A}\ \tau(f) 
=\int_{{\cal M}}{\rm Tr}_{\cal A}\ f\ ,\qquad
\int_{{\cal M}}{\rm Tr}_{\cal A}\ \pi_{k_y}(f)
=\int_{{\cal M}}{\rm Tr}_{\cal A}\ f\ .
\ee
It can be seen that the form of the FCS action is compatible with imposing
combinations of the following conditions:
\be (A,\wt A,B,\wt B; V, \wt V, U, \wt U)^\dagger=(-\wt A,-A,B,-\wt B;-\wt V,-V,-U,\wt U)\ ,\label{dagger}\ee
\be \tau(A,\wt A,B,\wt B; V,\wt V,U,\wt U)=(-\wt A,- A,B,-\wt B; -\wt V,- V,-U,\wt U)\ ,\label{taucond}\ee
\be
\pi_{k}(A,\wt A,B,\wt B; V,\wt V,U,\wt U)
= (A,\wt A,-B,-\wt B; V,\wt V,-U,-\wt U)\ .\label{pik}
\ee
Demanding only \eq{dagger} yields a model with a real action, that we shall
refer to as the \emph{augmented non-minimal bosonic model},
consisting of massless particles of all integer spins, arising
in the twisted-adjoint zero-form
\be \Phi:=\left(\tfrac12(1-\pi_{k})B\right) \star k\ ,\label{Phi}\ee
and an additional sector of non-propagating modes
arising in the adjoint zero-form $\frac12(1+\pi_{k})B$.
Imposing also \eq{taucond} leads to a model,
that we shall refer to as the \emph{augmented minimal
bosonic model}, consisting of massless particles of
even spins and additional non-propagating modes.
Applying the projection \eq{pik} to the two aforementioned
models removes the non-propagating modes,
leading to the non-minimal and minimal bosonic models,
respectively, with the equations of motion
\be dA+A\star A+\Phi\star \wt \Phi=0\ ,\qquad d\Phi+A\star \Phi-\Phi\star\pi(\wt A)=0\ ,\ee
\be d\wt A+\wt A\star \wt A+\pi(\wt \Phi\star \Phi)=0\ ,\qquad d\wt\Phi+\pi(\wt A)\star \wt \Phi-\wt \Phi\star  A=0\ ,\ee
where we recall from \eq{pibarpi} that $\pi=\pi_y\pi_z$ and we 
have defined
\be \wt\Phi:=k\star \left(\tfrac12(1-\pi_{k})\wt B\right)\ .\label{wtPhi}\ee
Focusing on the minimal bosonic model and going
to the basis defined in \eq{defW},
the reality conditions and other projections 
in \eq{dagger}--\eq{pik} read
\begin{eqnarray}
W^\dagger &=& -W\;, \qquad \tau(W) = - W\;,\qquad \pi_{k}(W)=W\ , \label{r1}\\[5pt]
K^\dagger &=& K\;, \qquad \tau(K) = K\; ,\qquad \pi_{k}(K)=K\ .\label{r2}
\end{eqnarray}
In terms of the bosonic higher spin algebra
\be hs(4;\Real) = \left\{\ P \in {\rm Aq}(2) \quad {\rm s.t.}\quad P^\dagger = - P\ ,
\quad \pi\bar{\pi}(P) = P\ \right\}\ ,\ee
which consists of the minimal bosonic  subalgebra
\be hs_0(4;\Real) = \left\{\ P \in hs(4;\Real)  \quad {\rm s.t.}\quad
\tau(P) = - P\ \right\}\ ,\ee
and the coset
\be hs_1(4;\Real) = hs(4;\Real)/hs_0(4;\Real)\ ,\ee
we see that the conditions in Eqs. \eq{r1}--\eq{r2} amount to
\be 
W\in \bigoplus_{r,\bar r=0,1} hs_0(4;\Real)\star (\k_y)^{\star r}\star
(\bar \k_{\bar y})^{\star \bar r}\ ,\ee
\be iK\in  \bigoplus_{r,\bar r=0,1} hs_1(4;\Real)\star (\k_y)^{\star r}\star
(\bar \k_{\bar y})^{\star \bar r}\ .
\ee
Thus, the $(\kappa_y,\bar \kappa_{\bar y})$-independent
sectors of $W$ and $K$, respectively, consist of the 
minimal bosonic higher spin gauge fields,
which are real gauge fields of even spin,
and purely imaginary gauge fields
with odd spin.
%

\section{Vasiliev's extended system versus FCS gauge theory}
\label{sec:Vasi}

In this section, we exhibit how closed and central terms arise 
as topological degrees of freedom $\wt B$, activated within 
a consistent truncation, and compare them to those of a recent 
proposal by Vasiliev 
containing dynamical closed and central elements, referred to 
as Lagrangian forms, which have been proposed as effective 
actions, without any reference to a path integral formulation.
We shall instead propose an AKSZ path integral formulation of 
the FCS model, along the lines of \cite{Boulanger:2011dd,Boulanger:2012bj}, 
that allows the higher spin invariants to be interpreted as 
contributions to an effective action.
In order to give a precise model, we shall also
need to specify a certain higher spin geometry \cite{Sezgin:2011hq}.

\subsection{Reduction of the dynamical two-form}

We expand the restriction of $\wt B$ to $\partial{\cal M}_9$
over globally defined central and closed elements in 
${\cal E}|_{\partial{\cal M}_9}$.
A basis for this space is given by\footnote{
However, on more general noncommutative base manifolds
such as ${\cal X}_4 \times {\cal Z}_4$ with ${\cal Z}_4=T_2 \times T_2$, 
one may also combine odd elements from ${\cal Z}_4$ with odd 
elements from ${\cal X}_4$.}
$J_{\cal X}^{(r)}\star (J)^{\star m}\star (\overline J)^{\star\bar m}$ 
with $m,\bar m=0,1$ where $J_{\cal X}^{(r)}$, $r=1,\dots,N$, is a 
basis for the even subspace of the
de Rham cohomology on ${\cal X}_4$, and 
\be
J:=-\tfrac{i}8 dz^\a dz_\a \,\kappa_z\star \kappa_y\star (k+\bar k)\ ,\qquad
\overline J:=J^\dagger\ ,
\label{J1}
\ee
where $j_z$ is defined in \eq{jz}.
Thus, inserting the Ansatz
\be
K=0\ ,\qquad 
\widetilde B = \sum_{n=0}^\infty\sum_{r;m,\bar m}   \wt b_{r;m,\bar m;n} \, 
J_{\cal X}^{(r)}  \star (J)^{\star m}\star (\overline J)^{\star\bar m} \star B^{\star n}\ ,
\label{E3bis}
\ee
where 
\be J_{\cal X}^{(0)}=1\ ,\qquad \widetilde b_{0;0,0;n}=0\ ,\qquad d\wt b_{r;m,\bar m;n}=0\ ,\ee
into the field equations \eq{E2} yields
\bea
& F_W -\sum_{n=0}^\infty\sum_{r;m,\bar m}   \wt b_{r;m,\bar m;n} \, 
J_{\cal X}^{(r)}  \star (J)^{\star m}\star (\overline J)^{\star\bar m} \star B^{\star (n+1)}=0\ ,&
\nonumber \\
& D_W B = 0\ .&
\label{DEE}
\eea
As shown in Section 4.2, globally defined configurations 
can be obtained by choosing a structure group and a set of 
transition functions.
The resulting geometries can be characterized by invariant 
functionals \cite{Sezgin:2011hq} whose values can be fixed
at the level of a partition function using Lagrange multipliers 
\cite{Boulanger:2012bj}.
Expressing the invariants in terms of $\wt b_{r;m,\bar m;n}$ 
and the remaining data, the former can be determined in 
terms the latter. 
Thus, the classical moduli of the dynamical two-form $\wt B$ 
is a set of integration constants determined by boundary
conditions imposed within the context of a globally
defined formulation.


\subsection{Comparison to duality extended Vasiliev system}


In order to compare with Vasiliev's recent modification of his 
original system \cite{Vasiliev:2015mka}, we truncate ${\cal W}$ to 
${\rm Aq}(2)$ 
and take $\wt b_{r;m,\bar m;n}$ to vanish except 
$\wt b_{0;1,0;n}$, $\wt b_{0;0,1;n}$, 
$\wt b_{0;1,1;n}$ and $\wt b_{r;0,0;n}$.
Defining
\be 
{\cal V}= \sum_{n=0}^\infty \wt b_{0;1,0;n} B^{\star(n+1)}\ ,
\qquad
\overline{\cal V} =  \sum_{n=0}^\infty \wt b_{0;0,1;n} B^{\star(n+1)}\ ,
\ee
\be 
{\cal U}_0= \sum_{n=0}^\infty \wt b_{0;1,1;n} B^{\star(n+1)}\ ,\qquad 
 {\cal U}_{1}=  \sum_{n=0}^\infty \wt b_{1;0,0;n} B^{\star(n+1)}\ ,\qquad
{\cal U}_2 =  \sum_{n=0}^\infty \wt b_{2;0,0;n} B^{\star(n+1)}\ ,\ee
and assuming that $J_{[2]}=J_{\cal X}^{(1)}$ and $J_{[4]}=J_{\cal X}^{(1)}$
are of degree two and four, respectively, the equations take the form
\bea
&F_W  -{\cal V} \star J + \overline {\cal V} \star {\overline J} 
+ {\cal U} _0\star J \star {\overline J} 
+  {\cal U} _1\star J_{[2]}+{\cal U} _2\star J_{[4]}=0\ ,&
\nonumber \\
&D_W B = 0\ .&
\label{EVS1}
\eea
On the other hand, Vasiliev's recently proposed extended system 
\cite{Vasiliev:2015mka}, adapted to our notation, is given by
\bea
&&F_W  -{\cal V} \star J + \overline {\cal V} \star {\overline J} 
+  {\cal U} _0\star J \star {\overline J} +g  J \star {\overline J}
+  {\cal L}_{[2]}+{\cal L}_{[4]}=0\ ,
\\[5pt]
&&D_W B = 0\ ,
\label{EVS2}
\eea
where ${\cal L}_{[2]}$ and ${\cal L}_{[4]}$ are two new dynamical 
fields, referred to as Lagrangian forms, given by globally defined
central and closed elements of degrees two and four, respectively.
As far as the local dynamics is concerned, the two systems
are equivalent in form degrees zero and one, since one can always 
choose a representative for ${\cal L}_{[2]}$ that vanishes in a given 
coordinate chart.
In higher form degrees, the duality extended Vasiliev system contains 
the term $g J\star \overline J$ and the Lagrangian forms, which are
not present in the FCS system\footnote{
Whether such coupling can be obtained either by expanding $B$
around a constant background value or allowing the dependence 
of $\wt B$ on $B$ to contain a simple pole, remains to be seen.}.
In \cite{Vasiliev:2015mka}, the integral $\oint {\cal L}_{[2]}$ has 
been interpreted as a black hole charge, as has been substantiated 
by its evaluation \cite{Vasiliev:2015mka} on the Didenko--Vasiliev 
black hole solution \cite{Didenko:2008va}.
As for the integral of ${\cal L}_{[4]}$ over spacetime, 
it has been proposed \cite{Vasiliev:2015mka} as the generating
functional of correlators within the context of holography
\footnote{Another proposal for the black hole entropy and generating 
functional of correlators in higher spin gravity has been made 
in \cite{Sezgin:2011hq}.}.
An important open 
problem in this  framework is how to account for loop corrections. 
It has been suggested that the quantum mechanical effects may emerge from 
classical dynamics in an infinite dimensional space that has enough room 
to describe all multiparticle states in the system \cite{Vasiliev:2012vf}.
If true, this would be a drastically new way of looking at quantum gravity.
The tests of these proposals remain to be seen. 

In our approach, we propose a path integral formulation  of the FCS model 
along the same lines as the AKSZ construction  of \cite{Boulanger:2012bj} 
within the geometric framework of \cite{Sezgin:2011hq}. In this approach, the 
terms proportional to the closed and central elements in \eq{EVS1}, which are similar
to the Lagrangian form terms in \eq{EVS2} but play a different role, as the computation of the 
effective action proceeds in this case by means of path integral quantization rules which necessarily involves the FCS action itself. The advantage of this approach is the availability of path integral formulation for quantization. We leave the computation of quantum effects to a future work but we shall outline below the role of certain topological invariants in the construction of the on-shell effective action.

\subsection{On-shell actions from topological invariants}

Starting from an AKSZ path integral on ${\cal M}_9 = 
[0,\infty[\, \times {\cal X}_4\times {\cal Z}_4$,
where all fields vanish at $\{\infty\}\times {\cal X}_4\times {\cal Z}_4$ and
in addition $P|_{\{0\}\times {\cal X}_4\times {\cal Z}_4}=0$,
as required by the Batalin--Vilkovisky master equation,
one finds that $S_{\rm H}$ vanishes on-shell\footnote{
It would be interesting to examine in more
detail how the fields could depend nontrivially
on the extra dimension, for which Eq. \eq{C&L} could
provide a useful tool.
For example, on a manifold with the topology
${\cal X}_4\times{\cal Z}_4\times [0,1]$, one could
examine interpolations between two inequivalent solutions
to the field equations \eq{E2} and the resulting
on-shell value for the action \eq{SH}.}.
Following \cite{Sezgin:2011hq}, one may generate 
an on-shell action by adding to $S_{\rm H}$ a globally defined boundary 
term $S_{\rm top}=\oint_{\partial{\cal M}_9} {\cal V}(X,dX)$,
whose total variation vanishes off-shell, 
\emph{i.e.} $S_{\rm top}$ is a topological invariant.
Assuming that $S_{\rm top}$ does not affect the boundary
condition on $P$ nor the equations of motion,
one may argue that the on-shell action
is given by $S_{\rm top}$.

Let us assume that $G$ is generated by a subalgebra 
of the algebra gauged by $(A,\wt A)$. Hence, splitting $A= \Gamma+E$ 
and $\wt A= \wt\Gamma +\wt E$ where $(\Gamma,\wt\Gamma)$
is the duality extended bundle connection and $(E,\wt E)$ a 
generalized soldering form, and taking ${\cal X}_{2p}\subset {\cal X}_4$ 
to be closed cycles of dimension $2p$ for $p=0,1,2$,
we consider\footnote{Instead of ${\cal X}_4$ one may
consider an open region ${\cal M}_4$ and subtract the
Chern--Deligne modified Chern--Simons form on $\partial{\cal M}_4\times {\cal Z}_4$
from $S_{\rm top}$ in order to obtain a well defined 
topological invariant. }
\be 
S_{\rm top}[\Gamma,\wt\Gamma]=\sum_{p=0}^2\sum_{n=1}^{p+2} \oint_{{\cal X}_{2p}\times
{\cal Z}_4} {\rm Tr}_{{{\cal W}}\otimes {\cal K}} \left(\a_{n,p} (F_\Gamma)^{\star n}+\wt \a_{n,p} (F_{\wt \Gamma})^{\star n}\right)\ ,
\ee
where $F_\Gamma=d\Gamma+\Gamma\star \Gamma$, $F_{\wt\Gamma}=
d{\wt\Gamma}+\wt\Gamma\star \wt\Gamma$ and $\a_{n,p}$ and $\wt \a_{n,p}$ are constants.
In the semi-classical limit, one thus has the partition function
\be Z_{\rm FCS}=\sum_{\rm saddles} {\cal N} e^{iS_{\rm top}}\ .\ee

In particular, for the diagonal embedding of the structure group defined by
\be \Gamma=\wt\Gamma = W\ ,\qquad E = -\wt E = -K\ ,\label{diag}\ee
one may take
\be 
S_{\rm top}[W,K]=\sum_{p=0}^2\sum_{n=1}^{p+2} \oint_{{\cal X}_{2p}\times
{\cal Z}_4} \left. \beta_{n,p}\left(\tfrac{d}{dt}\right){\rm Tr}_{{\cal W}
\otimes {\cal K}}  \left(F_{W_t}\right)^{\star n}\right\vert_{t=0}\ ,
\label{Stop}
\ee
where 
\be W_t= W+ t K\ ,\qquad F_{W_t}=F_W+ t D_W K+ t^2 K\star K\ ,\ee
and $\b_{n,p}$ are linear differential operators of order $(2n-1)$ in $d/dt$ with constant 
coefficients.
The on-shell value of $S_{\rm top}[W,K]$ is built out of integrals of
traces of $B\star \wt B$, $\wt B\star B$ and $K\star K$ 
forming a finite set of invariants provided that 
$(K,B,\wt B)$ and their gauge parameters belong to sections of the 
diagonal structure group.
For the diagonal structure group defined in \eq{diag},
the fields $(K,B,\wt B)$ and their gauge parameters $(\beta,
\eta,\wt \eta)$ belong to sections.
The observables are invariant
off shell under gauge transformations with parameter $\a$, and on shell
using parameters $(\b,\eta,\wt \eta)$.

\section{Linearized field equations}\label{sec:Boundary}

In this section, we shall assume that the structure group is
given by \eq{diag}, and linearize the duality extended 
FCS field equations around a vacuum for $\widetilde B$ 
given by a central and closed form $I$. 
As we shall see, the local degrees of freedom 
are indeed contained in the integration constant for the 
zero-form $B_{[0]}$.
Moreover, as the above choice of structure group implies 
that the gauge parameter $\widetilde \eta$ of $\widetilde B$ belongs to a 
globally defined section on ${\cal Z}_4$, some of the 
Killing symmetry parameters are not globally defined and 
the gauge symmetries of $\widetilde B$ cannot be used 
to gauge away its vacuum expectation value.
%

\subsection{Vacuum solution and Killing symmetries}

The theory on the boundary of ${\cal M}_9$ admits the vacuum solutions
\be \wt B^{(0)}=I\ ,\qquad W^{(0)}=L^{-1}\star dL\ ,\qquad
K^{(0)}=0\ ,\qquad B^{(0)}=0\ ,\label{vac}\ee
where $L$ is a gauge function (consisting of forms) and 
$I$ is a closed and central element on $\partial{\cal M}_9$.
The Killing symmetries obey
\bea
\delta B^{(0)}&\equiv&  D^{(0)}\eta^{(0)} =0\ ,
\label{k3}\\
\delta K^{(0)}&\equiv& D^{(0)}\beta^{(0)} =0\ ,
\label{k2}\\
\delta W^{(0)}&\equiv&  D^{(0)}\alpha^{(0)} + \tfrac12\{I,  \eta^{(0)}\}_\star =0\ ,
\label{k1}\\
\delta \wt B^{(0)}&\equiv& D^{(0)} {\widetilde\eta}^{(0)} -   \{I,
\beta^{(0)}\}_\star=0\ ,
\label{k4}
\eea
where $D^{(0)}\equiv D_{W^{(0)}}$.
Hence, 
\be
(\eta^{(0)},\beta^{(0)},\alpha^{(0)},\widetilde\eta^{(0)})
= L^{-1}\star (\eta^{(0)\prime}, \beta^{(0)\prime},\alpha^{(0)\prime},\widetilde\eta^{(0)\prime})\star L\ ,
\ee
where the primed fields obey Eqs. \eq{k3}--\eq{k4}
with $D^{(0)}$ replaced by $d$.

To solve \eq{k3}--\eq{k4}, it is useful to introduce the homotopy contractor
\be
\rho_v = \imath_v ({\cal L}_v)^{-1}=\imath_v \int_0^1 \frac{dt}t t^{{\cal L}_v}\ ,
\qquad {\cal L}_v=\{\imath_v, d\}\ ,\label{ho}
\ee
where $v$ is a vector field that generates a set of flow lines
emanating from a base point in $\partial{\cal M}_9$, which we shall
choose to be 
\be p_0=\{x_0\}\times \{\vec 0\}\in{\cal X}_4\times {\cal Z}_4\ .\ee
Thus, for $v=z^\a \partial_\a$ the contractor gives
\be
\rho_v f(Z,Y,dZ) = Z^\ua \frac{\partial}{\partial dZ^\ua} 
\int_0^1 dt \frac{1}{t} f(tZ,Y,tdZ)\ .
\ee
The contractor defined in \eq{ho} satisfies the relations 
\be \rho_v d \rho_v=\rho_v\ ,\qquad (\rho_v)^2=0\ .\ee
We shall assume that ${\cal L}_v$ has a well defined action 
on smooth symbols, such that
\be
f=\{\rho_v, d\}  f+\delta_{{\rm deg}(f),0} f|_{p_0}\ .
\ee
It follows that 
\be 
df=g\ ,\quad dg=0 \qquad\Longleftrightarrow\qquad
f = \rho_v g + f_{[0],0} + {d}\rho_v f \ ,\ee  
where $f_{[0],0}$ is a (zero-form) integration constant
and $\rho_v f$ is a gauge function.
Using these formula, the solutions of \eq{k3}--\eq{k4} are found to be 
\bea
\eta^{(0)\prime}&=& d\rho_v \eta^{(0)\prime}\ ,
\label{s1}\\
\beta^{(0)\prime}&=&\beta^{(0)\prime}_{[0]}+d\rho_v \beta^{(0)\prime}\ ,
\label{s2}\\
\alpha^{(0)\prime}&=&\alpha^{(0)\prime}_{[0]}+d\rho_v \alpha^{(0)\prime}
+\tfrac12 (d\rho_v -1)\{I,\rho_v\eta^{(0)\prime}\}_\star\ ,
\label{s3}\\
\widetilde\eta^{(0)\prime}&=& d\rho_v \widetilde\eta^{(0)\prime}
-(d\rho_v-1) \left(\{\rho_v I,\beta^{(0)\prime}_{[0]}\}_\star
+\{I,\rho_v\beta^{(0)\prime}\}_\star\right)\ ,\qquad
\label{tildeeta0}
\eea
where $\rho_v(\eta^{(0)\prime},\a^{(0)\prime},\b^{(0)\prime},\wt\eta^{(0)\prime})$
are gauge functions, $\alpha^{(0)\prime}_{[0]}$ and $ \beta^{(0)\prime}_{[0]} $
are integration constants. In obtaining \eq{s3} we have used \eq{s1} and in obtaining 
\eq{tildeeta0} we have used \eq{s2} and the fact that $\rho_vd\rho_v=\rho_v$. 

In order to describe Vasiliev's phase of the theory, we assume that
\be I=J_{\cal X}+e^{i\theta_0}J-e^{-i\theta_0}\overline J\ ,\ee
where $J_{\cal X}$ is a closed a central element  
on ${\cal X}_4$, $J$ and $\overline J$ are the 
closed and central elements on ${\cal Z}_4$ defined 
in \eq{J1}, and $\theta_0$ is an arbitrary real constant.
The restriction of \eq{tildeeta0} to form degree one reads
\be 
\widetilde\eta^{(0)\prime}_{[1]}={\rm d}\rho_v \widetilde\eta^{(0)\prime}_{[1]}
+\{\rho_v I,\beta^{(0)\prime}_{[0]}\}_\star\ ,
\label{eta1}
\ee
which furnishes a decomposition of $\widetilde\eta^{(0)\prime}_{[1]}$ 
into two linearly independent terms, as the first term is $d$-exact 
and the second term is in the co-kernel of $d$.
Since  $\widetilde\eta^{(0)\prime}_{[1]}$ must be globally 
defined on ${\cal Z}_4$, it follows that the two terms 
in \eq{eta1} must be separately globally defined 
on ${\cal Z}_4$.
As for the $d$-exact term, this constrains  
$\rho_v \widetilde\eta^{(0)\prime}_{[1]}$, while 
the second term can only be globally defined if 
it vanishes, \emph{i.e.} if  
\be \beta^{(0)\prime}_{[0]}=0\ ,\ee
since $\rho_v J$ cannot be globally defined on ${\cal Z}_4$,
as this would contradict the fact that 
$\int_{{\cal Z}_4} J\star\overline J$ is non-vanishing.
Hence, the Killing parameters for the duality unextended system are
\be
\alpha^{(0)}_{[0]}=L^{-1}\star \alpha^{(0)\prime}_{[0]}\star L\ ,
\qquad \beta^{(0)}_{[0]}=0\ ,\qquad
\widetilde\eta^{(0)}_{[1]}=L^{-1}\star {d}(\rho_v \widetilde\eta^{(0)\prime}_{[1]})\star L
\ ,\label{K3}
\ee
where the $(\kappa_y,\bar \kappa_{\bar y})$-independent 
subsector of $\alpha^{(0)}_{[0]}$ coincides with the Killing
symmetries of Vasiliev's extended system.
We leave the analysis of the Killing parameters 
in higher form degree for future work.

\subsection{Linearized fluctuations}

We expand the fluctuations in the boundary fields as
\be
(W-W^{(0)},B,K,\wt B-\wt B^{(0)})=\sum_{n\geqslant 1}
( W^{(n)},B^{(n)},K^{(n)},\wt B^{(n)})\ .
\ee
At the first order, the equations of motion \eq{eom2} read
\be
D^{(0)} W^{(1)}-\tfrac12\{I, B^{(1)}\}_\star=0\ ,\qquad
D^{(0)} K^{(1)} =0\ , \ee
\be D^{(0)} B^{(1)}=0\ ,\qquad D^{(0)} \wt B^{(1)}
+\{I, K^{(1)}\}_\star=0\ .
\label{abe}
\ee
The abelian gauge transformations following from \eq{dt4} are given by
\be
\delta W^{(1)}=D^{(0)} \alpha^{(1)}+\tfrac12\{I, \eta^{(1)}\}_\star\ ,\qquad
\delta B^{(1)}=D^{(0)} \eta^{(1)}\ ,\ee\be
\delta K^{(1)}=D^{(0)} \beta^{(1)} \ ,\qquad
\delta \wt B^{(1)}=D^{(0)} \wt\eta^{(1)} 
- \{I, \beta^{(1)}\}_\star\ .
\ee
%
%
%
Using the homotopy contractor, the linearized fluctuations
can be expressed in terms of the gauge function and an
integration constant $B^{(1)'}_{[0]}$ as
\be
(W^{(1)},B^{(1)},K^{(1)},\wt B^{(1)})=L^{-1}\star (W^{(1)\prime},B^{(1)\prime},K^{(1)\prime},\wt B^{(1)\prime})\star L\ ,
\ee
where
\bea
B^{(1)\prime}&=&B^{(1)\prime}_{[0]}+{\rm d}\rho_v B^{(1)\prime}\ ,\\[5pt]
W^{(1)\prime}&=&
d\rho_v W^{(1)\prime}-\tfrac12 (d\rho_v-1) \left(\{\rho_v I,
B_{[0]}^{(1)\prime} \}_\star +\{I ,\rho_v B^{(1)\prime}\}_\star\right)\ ,\\[5pt] 
K^{(1)\prime}&=&d\rho_v K^{(1)\prime}\ ,\\[5pt]
\wt B^{(1)\prime}
&=&d\rho_v\wt B^{(1)\prime}+(d\rho_v-1)\{I,\rho_v K^{(1)\prime}\}_\star\ .
\eea
The local degrees of freedom of the system are contained
in the zero-form integration constant $B^{(1)\prime}_{[0]}$.
The connection $W^{(1)}_{[1]}$ consists of a pure gauge 
solution, as its gauge function and gauge parameter belong to the 
same spaces, plus a a set particular solutions that carrying the 
aforementioned local massless degrees of freedom.

The fields $K^{(1)}$ and $\wt B^{(1)\prime}$, on the other hand, 
may introduce new topological degrees of freedom arising in
cohomological spaces given by spaces of gauge functions over
the spaces of gauge parameters. 
In particular, $\wt B^{(1)\prime}_{[2]}$ contains moduli associated 
to the gauge function $\rho_v \wt B^{(1)\prime}_{[2]}=
\rho_v (e^{i\theta_0}J-e^{-i\theta_0}\bar J)$, 
as $d\rho_v (e^{i\theta_0}J-e^{-i\theta_0}\bar J)=
e^{i\theta_0}J-e^{-i\theta_0}\bar J$ belongs to an admissible section 
for $\wt B^{(1)\prime}_{[2]}$ while 
$\rho_v (e^{i\theta_0}J-e^{-i\theta_0}\bar J)$ does not belong 
to an admissible section for $\wt \eta^{(1)}$.
In order to exhibit in more detail the nature of
the topological degrees of freedom in $\wt B$,
we consider the representative
\be 
\rho_v (e^{i\theta_0}J-e^{-i\theta_0}\bar J)=
(e^{i\theta_0}v'\star \kappa_y-e^{-i\theta_0}\bar v'\star 
\bar \kappa_{\bar y})\star k\star \Pi^+_{k,\bar k}\ ,\ee
obtained by taking $v=z^\a \partial_\a$.
It follows that $v'$ is the solution to 
\be dv'= j_z\ ,\qquad \imath_v v'=0\ ,\ee
on the interior of ${\cal R}^\Real_4$ given by
\be v'= \frac{dz^\a z_\a}{z^+ z^-}\ ,\ee
where $z_\pm$ are defined by splitting $z^\a$ such that $[z^-,z^+]_\star=1.$
This element does not belong to $\Omega({\cal Z}_4)$, 
as it does not fall off fast enough at infinity.
Thus, the moduli of $\wt B^{(1)\prime}_{[2]}$ 
can be associated to modes that blow up at 
infinity, \emph{i.e.} at the commutative point 
of ${\cal Z}_4$.

At higher orders in perturbative expansion,
the moduli of $\wt B$ will generate the interaction
terms contained in the Ansatz \eq{E3bis}.
On a more general base manifold ${\cal M}_9$, 
it follows from the fact that $K$, $B$ 
and $\widetilde B$ belong to sections of the 
structure group that they can contain topological 
degrees of freedom given by matching
elements in the de Rham cohomology, whose r\^ole 
remains to be investigated further.

The above linearization suffices to show that the perturbative degrees of freedom
of the system are contained in the initial data for the Weyl zero-form. 
However, in order obtain Fronsdal field equations one has 
switch from Weyl order to normal order and perform a change of
gauge in order to make direct contact with Vasiliev's 
original perturbative expansion (in which $z^\a A_\a=0$ in normal order), 
which complies with the Central On Mass Shell Theorem (COMST).
It is important that despite the fact that the FCS model 
is formulated in the Weyl order, for reasons explained 
in Section 4, its physical spectrum agrees with the 
Vasiliev theory, and hence its perturbative expansion 
should obey the COMST as well.
Although a naive transformation of the perturbatively defined
master fields from normal to Weyl order 
is known to produce singularities \cite{Vasiliev:2015wma}
\footnote{For a general discussion of ordering schemes 
and maps between them, see \emph{e.g.} \cite{zachos2005quantum}.},
the FCS master fields belong to an extended class of symbols, 
including inner Klein operators, which yields a well-defined 
perturbation theory in a specific holomorphic gauge (defined
by $z^\a A_\a=0$ in Weyl order). 
Indeed, working with definite boundary conditions 
(corresponding to generalized Type D solutions \cite{Iazeolla:2007wt}), 
the resulting linearized fields can be mapped to Vasiliev's basis.
We plan to examine whether this remains the case for more general boundary
conditions and to higher orders in the perturbative expansion.

\section{Conclusions}\label{sec:Conclusions}

We have extended Vasiliev's higher spin gravity in four dimensions
by elevating the rigid two-form in twistor space to a
dynamical master field, and introducing a new one-form master
field to achieve Cartan integrability, thereby enlarging the higher spin
symmetry. 
We have also introduced suitable Lagrange multipliers 
such that the resulting complete field content fits 
into a tensor product of the higher spin algebra and an eight
dimensional $\mathbb Z_2$-graded Frobenius algebra 
with bi-linear form given by a trace operation.
Furthermore, in order provide an action principle suitable
for AKSZ quantization, we have introduced an auxiliary dimension 
by taking the master fields to live in a direct product
of a five-dimensional open manifold, with boundary containing
spacetime, and a closed version of the Vasiliev's noncommutative
twistor $Z$-space.
Assembling all fields into a superconnection valued
in the aforementioned associative algebra, we have
found a natural generalization of the Chern--Simons
action to higher dimensions, that remains with 
quadratic kinetic terms, which we refer to as
the Frobenius--Chern--Simons (FCS) action.
The variational principle and gauge invariance
is ensured by imposing appropriate boundary
conditions and working with a suitable class
of functions on the twistor $Z$-space.

An important ingredient of the FCS model 
is taking the master fields to consist of 
sums of appropriate even and odd forms 
in accordance with the dimension of the 
base manifold and requiring absence of 
zero-form constraints on-shell.
The procedure of adding higher forms that 
are sourced by closed and central terms 
on-shell was proposed in \cite{Boulanger:2011dd},
where it was called duality extension, and it
has important consequences as explained already
in the introduction.
The results presented in Section \ref{sec:Vasi} show explicitly
the manner in which these higher forms give rise 
to novel modifications of standard Vasiliev equations.
A subset of these interactions have been recently pointed out in
\cite{Vasiliev:2015mka} at the level of field equations.
It is important to note that while the 
duality extension does not add any local 
degrees of freedom, they effect the interactions 
in crucial ways. 
It would be interesting to study the weak field expansions 
in presence the attendant higher form.

An advantage of the FCS formulation
of higher spin gravity is that it 
enlarges the symmetries present in the
standard Vasiliev system without adding
any new perturbative degrees of freedom.
This implies a drastic restriction of 
the topological terms that can be added 
to the covariant Hamiltonian action 
without ruining its salient features (also in the presence 
of sources which may be of relevance to the 
embedding of the FCS model into string theory),
to produce an on-shell action,
as discussed in Section 6.
Based on sample calculations, we expect the 
perturbative expansion of the resulting on-shell 
action to yield higher spin amplitudes reproducing
proposed conformal field theory data.

Having summarized our results, we turn to  
open problems and future directions.
An outstanding problem in higher spin gravity
is the computation of the free energy.
At the leading order, it has been hindered 
so far by the lack of an action principle, 
while the one-loop computation has been performed
\cite{Giombi:2013fka} under the assumption 
that the action has an expansion about the anti-de 
Sitter vacuum in terms of Fronsdal fields.  
Whether the same one-loop contribution to 
the free energy will ultimately emerge 
from the FCS action is far from clear,
and remains to be seen.

It is often stated that covariant Hamiltonian
actions of the type constructed in \cite{Boulanger:2011dd}
and here are not acceptable because they do not contain 
the quadratic Fronsdal action. 
However, these arguments need to be scrutinized.
To begin with, the covariant Hamiltonian actions do give 
the fully nonlinear 
field equations which describe the propagating fields and 
their interactions at the classical level.
Moreover, these actions can be quantized 
following the AKSZ methods employed already 
in \cite{Boulanger:2012bj}
\footnote{It would also be interesting to connect the
covariant Hamiltonian actions with the
Kazinski--Lykhanovich--Sharapov approach to actions for
non-Lagrangian theories \cite{Kazinski:2005eb}
which also relies on the introduction of an auxiliary dimensions.},
leading to bulk Feynman diagrams of a different type
than those resulting from a would-be nonlinear completion 
of the Fronsdal action.
So far, loop computations with external fields have not been computed in 
either approach. Even though the Fronsdal program approach may be 
forbiddingly difficult to achieve for $N$-point functions with $N>5$,
it would be interesting to compare the final results, at least for one-loop 
corrections to N=2,3,4 point functions with those computed from the FCS model. 
In this context, we note that both in Fronsdal program approach as well as the AKSZ approach
the classical moduli spaces overlap on initial data but differ at nonlinear level viewed as classical solutions. However, if the uniqueness arguments for higher spin gravities can be strengthened, one may expect a nontrivial duality relation between the two formulations, both of which, in turn  should be dual to the same CFT on the boundary of $AdS_4$. 

Regarding the vacuum energy, we can make the following observations in the case 
of the FCS model. Following the AKSZ procedure \cite{Alexandrov:1995kv},
the BV master equation for the path integral measure 
$\left\langle\cdot\right\rangle_{\rm H}$ requires the 
action functional $S_{\rm H}$ to be differentiable.
This requires that $P|_{\partial {\cal M}_9}=0$ {\it off-shell}
(such that $S_{\rm H}=S_{\rm FCS}$).
Dropping the $P^{\star3}$ terms and working in Weyl order, 
it follows from \eq{fstarg} that
\be S^{\rm free}_{\rm H} = \int_{{\cal M}_9} {\rm Tr}_{\cal A} \,P {d}X=
\sum_{p=0}^{{\rm dim}{\cal M}-1}
\sum_{\tau_p}\int_{{\cal M}_9} P^{[9-p]}_{\tau_p} {d}X^{\tau_p}_{[p]}\ ,
\label{abelianpform}\ee
where for each $p$, the sum over $\tau_p$ runs
over an infinite set of component fields.
For each $\tau_p$, the contribution to the partition function ${\cal Z}$
from the resulting abelian $p$-form
system \cite{Schwarz:1978cn,Schwarz:1984wk,Horowitz:1989km,Wu:1990ci}
is given by an integral over a space of zero modes $\mu$
with integrand given by the functional determinants
of the non-zero modes.
The latter combine into the topological invariant \cite{Wu:1990ci}
\be {\cal Z}_{\tau_p}(\mu) = ({\rm Tor}({\cal M}))^{(-1)^{p+1}}\ ,\label{tor}\ee
where ${\rm Tor}({\cal M})$ is the Ray-Singer analytical torsion
for manifolds with boundaries.
Thus, modulo zero-mode issues, rearranging the products 
in ${\cal Z}=\int d\mu \prod_p \prod_{\tau_p} {\cal Z}_{\tau_p}(\mu)$,
shows that there are cancellations between
even and odd forms creating a balance in the augmented models 
but not in their $\pi_k$-projected counterparts, as defined in Section 2.8.
Therefore the augmented models appear to be
well-defined at one-loop order,
and the FCS model may provide a viable framework
for semi-classical calculations, as outlined in Section 5.
In \cite{Engquist:2005yt}, based on properties
of solitons of the Nambu--Goto action and 
discretization of the spatial coordinate of
the worldsheet, it was argued that tensionless 
closed strings in anti-de Sitter spacetime
disintegrate into more fundamental objects 
made up out of totally symmetric 
multi-singleton states\footnote{According to
this proposal, the Hagedorn phenomenon 
marks a phase transition and not a maximal
temperature.}.
In particular, the two-singleton sector was 
mapped to a topological open string with 
amplitudes given by traces over Weyl algebras.
Later, in \cite{Colombo:2012jx,Didenko:2012tv}
traces of the same type, derived from the zero-form 
charges in \eqref{I2n} were shown to reproduce 
conformal field theory correlation functions.
Moreover, the quantization of Poisson manifolds 
\cite{Kontsevich:1997vb} naturally leads to
two-dimensional sigma models \cite{Cattaneo:1999fm,Cattaneo:2001ys}.
This framework extends to differential Poisson algebras  \cite{Chu:1997ik,Beggs:2003ne,McCurdy:2008ew,McCurdy:2009xz,Arias:2015wha},
leading to models \cite{Bonezzi:2015lfa} in which
the de Rham differential in target space is gauged.
Given the fact that topological open string field 
theory naturally leads to Frobenious algebras 
\cite{Gaberdiel:1997ia} and the master fields
with gapless expansions in form degrees, 
we would like to propose that such a model
contains the duality extended FCS model as
a classically consistent truncation.

\paragraph{Acknowledgements:}

We thank Cesar Arias, Massimo Bianchi, Pierre Bieliavski, Roberto Bonezzi, 
Alex Gomez--Torres, Rodrigo Olea, Valentin Ovsienko, Patricia Ritter, Andy Royston, 
Augusto Sagnotti, Eugene Skvortsov, Mauricio Valenzuela, 
Brenno Vallilo, Misha Vasiliev and Andy Waldron for discussions.
N.B. wants to thank the IH\'ES for hospitality.
He is F.R.S.-FNRS Research Associate (Belgium) and 
his work was supported in parts by an
ARC contract No. AUWB-2010-10/15-UMONS-1.
The work of E.S. is supported in part by NSF grant PHY-1214344.
P.S. is grateful to Texas A\&M University, UMONS and 
NITheP (Stellenbosch) for hospitality during various stages of this work.
The work of P.S. is supported by Fondecyt Regular grant N$^{\rm o}$
1140296 and Conicyt grant DPI 20140115.

\providecommand{\href}[2]{#2}\begingroup\raggedright\endgroup


\begin{thebibliography}{10}

\bibitem{Vasiliev:1990en}
M.~A. Vasiliev, ``{Consistent equation for interacting gauge fields of all
  spins in (3+1)-dimensions},'' {\em Phys. Lett.} {\bf B243} (1990)
378--382.

\bibitem{Vasiliev:1992av}
M.~A. Vasiliev, ``{More on equations of motion for interacting massless fields
  of all spins in (3+1)-dimensions},'' {\em Phys. Lett.} {\bf B285} (1992)
225--234.

\bibitem{Vasiliev:2003ev}
M.~A. Vasiliev, ``{Nonlinear equations for symmetric massless higher spin
  fields in (A)dS(d)},'' {\em Phys. Lett.} {\bf B567} (2003) 139--151,
\href{http://arXiv.org/abs/hep-th/0304049}{{\tt hep-th/0304049}}.

\bibitem{Fronsdal:1978rb}
C.~Fronsdal, ``{Massless Fields with Integer Spin},'' {\em Phys. Rev.} {\bf
  D18} (1978)
3624.

\bibitem{Vasiliev:1986td}
M.~A. Vasiliev, ``{Free massless fields of arbitrary spin in the de Sitter
  space and initial data for a higher spin superalgebra},'' {\em Fortsch.Phys.}
  {\bf 35} (1987)
741--770.

\bibitem{Lopatin:1987hz}
V.~Lopatin and M.~A. Vasiliev, ``Free massless bosonic fields of arbitrary spin
  in d-dimensional de sitter space,'' {\em Mod.Phys.Lett.} {\bf A3} (1988) 257.

\bibitem{Vasiliev:2001wa}
M.~A. Vasiliev, ``{Cubic interactions of bosonic higher spin gauge fields in
  AdS(5)},'' {\em Nucl. Phys.} {\bf B616} (2001) 106--162,
\href{http://arXiv.org/abs/hep-th/0106200}{{\tt hep-th/0106200}}.

\bibitem{Fradkin:1986qy}
E.~S. Fradkin and M.~A. Vasiliev, ``{Cubic Interaction in Extended Theories of
  Massless Higher Spin Fields},'' {\em Nucl. Phys.} {\bf B291} (1987)
141.

\bibitem{Vasilev:2011xf}
M.~Vasiliev, ``{Cubic Vertices for Symmetric Higher-Spin Gauge Fields in
  $(A)dS_d$},'' {\em Nucl.Phys.} {\bf B862} (2012) 341--408,
\href{http://arXiv.org/abs/1108.5921}{{\tt 1108.5921}}.

\bibitem{Buchbinder:2006eq}
I.~L. Buchbinder, A.~Fotopoulos, A.~C. Petkou, and M.~Tsulaia, ``{Constructing
  the cubic interaction vertex of higher spin gauge fields},'' {\em Phys. Rev.}
  {\bf D74} (2006) 105018,
\href{http://arXiv.org/abs/hep-th/0609082}{{\tt hep-th/0609082}}.

\bibitem{Metsaev:2006ui}
R.~R. Metsaev, ``{Gravitational and higher-derivative interactions of massive
  spin 5/2 field in (A)dS space},'' {\em Phys. Rev.} {\bf D77} (2008) 025032,
\href{http://arXiv.org/abs/hep-th/0612279}{{\tt hep-th/0612279}}.

\bibitem{Fotopoulos:2007yq}
A.~Fotopoulos, N.~Irges, A.~C. Petkou, and M.~Tsulaia, ``{Higher-Spin Gauge
  Fields Interacting with Scalars: The Lagrangian Cubic Vertex},'' {\em JHEP}
  {\bf 10} (2007) 021,
\href{http://arXiv.org/abs/0708.1399}{{\tt 0708.1399}}.

\bibitem{Zinoviev:2008ck}
Y.~M. Zinoviev, ``{On spin 3 interacting with gravity},'' {\em Class. Quant.
  Grav.} {\bf 26} (2009) 035022,
\href{http://arXiv.org/abs/0805.2226}{{\tt 0805.2226}}.

\bibitem{Boulanger:2008tg}
N.~Boulanger, S.~Leclercq, and P.~Sundell, ``{On The Uniqueness of Minimal
  Coupling in Higher-Spin Gauge Theory},'' {\em JHEP} {\bf 0808} (2008) 056,
\href{http://arXiv.org/abs/0805.2764}{{\tt 0805.2764}}.

\bibitem{Boulanger:2011qt}
N.~Boulanger, E.~Skvortsov, and Y.~Zinoviev, ``{Gravitational cubic
  interactions for a simple mixed-symmetry gauge field in AdS and flat
  backgrounds},'' {\em J.Phys.A} {\bf A44} (2011) 415403,
  \href{http://arXiv.org/abs/1107.1872}{{\tt 1107.1872}}.

\bibitem{Joung:2012hz}
E.~Joung, L.~Lopez, and M.~Taronna, ``{Generating functions of
  (partially-)massless higher-spin cubic interactions},'' {\em JHEP} {\bf 1301}
  (2013) 168,
\href{http://arXiv.org/abs/1211.5912}{{\tt 1211.5912}}.

\bibitem{Boulanger:2012dx}
N.~Boulanger, D.~Ponomarev, and E.~Skvortsov, ``{Non-abelian cubic vertices for
  higher-spin fields in anti-de Sitter space},'' {\em JHEP} {\bf 1305} (2013)
  008,
\href{http://arXiv.org/abs/1211.6979}{{\tt 1211.6979}}.

\bibitem{Joung:2013nma}
E.~Joung and M.~Taronna, ``{Cubic-interaction-induced deformations of
  higher-spin symmetries},'' {\em JHEP} {\bf 1403} (2014) 103,
\href{http://arXiv.org/abs/1311.0242}{{\tt 1311.0242}}.

\bibitem{Metsaev:1991mt}
R.~R. Metsaev, ``{Poincare invariant dynamics of massless higher spins: Fourth
  order analysis on mass shell},'' {\em Mod. Phys. Lett.} {\bf A6} (1991)
359--367.

\bibitem{Taronna:2011kt}
M.~Taronna, ``{Higher-Spin Interactions: four-point functions and beyond},''
  {\em JHEP} {\bf 1204} (2012) 029,
\href{http://arXiv.org/abs/1107.5843}{{\tt 1107.5843}}.

\bibitem{Dempster:2012vw}
P.~Dempster and M.~Tsulaia, ``{On the Structure of Quartic Vertices for
  Massless Higher Spin Fields on Minkowski Background},'' {\em Nucl.Phys.} {\bf
  B865} (2012) 353--375,
\href{http://arXiv.org/abs/1203.5597}{{\tt 1203.5597}}.

\bibitem{Bekaert:2010hw}
X.~Bekaert, N.~Boulanger, and P.~A. Sundell, ``How higher-spin gravity
  surpasses the spin-two barrier,'' {\em Rev. Mod. Phys.} {\bf 84} (Jul, 2012)
  987--1009,
\href{http://arXiv.org/abs/1007.0435}{{\tt 1007.0435}}.

\bibitem{Alexandrov:1995kv}
M.~Alexandrov, M.~Kontsevich, A.~Schwartz, and O.~Zaboronsky, ``{The Geometry
  of the master equation and topological quantum field theory},'' {\em Int. J.
  Mod. Phys.} {\bf A12} (1997) 1405--1430,
\href{http://arXiv.org/abs/hep-th/9502010}{{\tt hep-th/9502010}}.

\bibitem{Boulanger:2012bj}
N.~Boulanger, N.~Colombo, and P.~Sundell, ``{A minimal BV action for Vasiliev's
  four-dimensional higher spin gravity},'' {\em JHEP} {\bf 1210} (2012) 043,
\href{http://arXiv.org/abs/1205.3339}{{\tt 1205.3339}}.

\bibitem{Boulanger:2011dd}
N.~Boulanger and P.~Sundell, ``{An action principle for Vasiliev's
  four-dimensional higher-spin gravity},'' {\em J.Phys.} {\bf A44} (2011)
  495402,
\href{http://arXiv.org/abs/1102.2219}{{\tt 1102.2219}}.

\bibitem{Vasiliev:1988sa}
M.~A. Vasiliev, ``{Consistent equations for interacting massless fields of all
  spins in the first order in curvatures},'' {\em Annals Phys.} {\bf 190}
  (1989)
59--106.

\bibitem{Doroud:2011xs}
N.~Doroud and L.~Smolin, ``{An Action for higher spin gauge theory in four
  dimensions},'' \href{http://arXiv.org/abs/1102.3297}{{\tt 1102.3297}}.

\bibitem{Sezgin:2011hq}
E.~Sezgin and P.~Sundell, ``{Geometry and Observables in Vasiliev's Higher Spin
  Gravity},'' {\em JHEP} {\bf 07} (2012) 121,
\href{http://arXiv.org/abs/1103.2360}{{\tt 1103.2360}}.

\bibitem{Colombo:2010fu}
N.~Colombo and P.~Sundell, ``{Twistor space observables and quasi-amplitudes in
  4D higher spin gravity},'' {\em JHEP} {\bf 1111} (2011) 042,
\href{http://arXiv.org/abs/1012.0813}{{\tt 1012.0813}}.

\bibitem{Colombo:2012jx}
N.~Colombo and P.~Sundell, ``{Higher Spin Gravity Amplitudes From Zero-form
  Charges},''
\href{http://arXiv.org/abs/1208.3880}{{\tt 1208.3880}}.

\bibitem{Didenko:2012tv}
V.~Didenko and E.~Skvortsov, ``{Exact higher-spin symmetry in CFT: all
  correlators in unbroken Vasiliev theory},'' {\em JHEP} {\bf 1304} (2013) 158,
\href{http://arXiv.org/abs/1210.7963}{{\tt 1210.7963}}.

\bibitem{Engquist:2005yt}
J.~Engquist and P.~Sundell, ``{Brane partons and singleton strings},'' {\em
  Nucl. Phys.} {\bf B752} (2006) 206--279,
\href{http://arXiv.org/abs/hep-th/0508124}{{\tt hep-th/0508124}}.

\bibitem{Arias:2015wha}
C.~Arias, N.~Boulanger, P.~Sundell, and A.~Torres-Gomez, ``{2D sigma models and
  differential Poisson algebras},'' {\em JHEP} {\bf 08} (2015) 095,
\href{http://arXiv.org/abs/1503.05625}{{\tt 1503.05625}}.

\bibitem{Bekaert:2015tva}
X.~Bekaert, J.~Erdmenger, D.~Ponomarev, and C.~Sleight, ``{Quartic AdS
  Interactions in Higher-Spin Gravity from Conformal Field Theory},''
\href{http://arXiv.org/abs/1508.04292}{{\tt 1508.04292}}.

\bibitem{Vasiliev:2015mka}
M.~Vasiliev, ``{Invariant Functionals in Higher-Spin Theory},''
\href{http://arXiv.org/abs/1504.07289}{{\tt 1504.07289}}.

\bibitem{Sezgin:2003pt}
E.~Sezgin and P.~Sundell, ``{Holography in 4D (super) higher spin theories and
  a test via cubic scalar couplings},'' {\em JHEP} {\bf 0507} (2005) 044,
\href{http://arXiv.org/abs/hep-th/0305040}{{\tt hep-th/0305040}}.

\bibitem{Engquist:2002gy}
J.~Engquist, E.~Sezgin, and P.~Sundell, ``{Superspace formulation of 4D higher
  spin gauge theory},'' {\em Nucl. Phys.} {\bf B664} (2003) 439--456,
\href{http://arXiv.org/abs/hep-th/0211113}{{\tt hep-th/0211113}}.

\bibitem{Sezgin:2005pv}
E.~Sezgin and P.~Sundell, ``{An exact solution of 4D higher-spin gauge
  theory},'' {\em Nucl. Phys.} {\bf B762} (2007) 1--37,
\href{http://arXiv.org/abs/hep-th/0508158}{{\tt hep-th/0508158}}.

\bibitem{Iazeolla:2011cb}
C.~Iazeolla and P.~Sundell, ``{Families of exact solutions to Vasiliev's 4D
  equations with spherical, cylindrical and biaxial symmetry},'' {\em JHEP}
  {\bf 1112} (2011) 084,
\href{http://arXiv.org/abs/1107.1217}{{\tt 1107.1217}}.

\bibitem{Batalin:1981jr}
I.~Batalin and G.~Vilkovisky, ``{Gauge Algebra and Quantization},'' {\em
  Phys.Lett.} {\bf B102} (1981) 27--31.

\bibitem{Batalin:1984jr}
I.~Batalin and G.~Vilkovisky, ``{Quantization of Gauge Theories with Linearly
  Dependent Generators},'' {\em Phys.Rev.} {\bf D28} (1983) 2567--2582.

\bibitem{Cattaneo:2001ys}
A.~S. Cattaneo and G.~Felder, ``{On the AKSZ formulation of the Poisson sigma
  model},'' {\em Lett. Math. Phys.} {\bf 56} (2001) 163--179,
\href{http://arXiv.org/abs/math/0102108}{{\tt math/0102108}}.

\bibitem{Ikeda:2001fq}
N.~Ikeda, ``{Deformation of BF theories, topological open membrane and a
  generalization of the star deformation},'' {\em JHEP} {\bf 07} (2001) 037,
\href{http://arXiv.org/abs/hep-th/0105286}{{\tt hep-th/0105286}}.

\bibitem{Hofman:2002jz}
C.~Hofman and J.-S. Park, ``{BV quantization of topological open membranes},''
  {\em Commun. Math. Phys.} {\bf 249} (2004) 249--271,
\href{http://arXiv.org/abs/hep-th/0209214}{{\tt hep-th/0209214}}.

\bibitem{Roytenberg:2006qz}
D.~Roytenberg, ``{AKSZ-BV formalism and Courant algebroid-induced topological
  field theories},'' {\em Lett. Math. Phys.} {\bf 79} (2007) 143--159,
\href{http://arXiv.org/abs/hep-th/0608150}{{\tt hep-th/0608150}}.

\bibitem{Barnich:2009jy}
G.~Barnich and M.~Grigoriev, ``{A Poincare lemma for sigma models of AKSZ
  type},'' {\em J.Geom.Phys.} {\bf 61} (2011) 663--674,
\href{http://arXiv.org/abs/0905.0547}{{\tt 0905.0547}}.

\bibitem{Alkalaev:2013hta}
K.~Alkalaev and M.~Grigoriev, ``{Frame-like Lagrangians and presymplectic
  AKSZ-type sigma models},'' {\em Int.J.Mod.Phys.} {\bf A29} (2014), no.~18,
  1450103,
\href{http://arXiv.org/abs/1312.5296}{{\tt 1312.5296}}.

\bibitem{Vasiliev:2007yc}
M.~Vasiliev, ``{On Conformal, SL(4,R) and Sp(8,R) Symmetries of 4d Massless
  Fields},'' {\em Nucl.Phys.} {\bf B793} (2008) 469--526,
  \href{http://arXiv.org/abs/0707.1085}{{\tt 0707.1085}}.

\bibitem{Quillen198589}
D.~Quillen, ``Superconnections and the chern character,'' {\em Topology} {\bf
  24} (1985), no.~1, 89 -- 95.

\bibitem{Bonezzi:2015lfa}
R.~Bonezzi, P.~Sundell, and A.~Torres-Gomez, ``{2D Poisson Sigma Models with
  Gauged Vectorial Supersymmetry},'' {\em JHEP} {\bf 08} (2015) 047,
\href{http://arXiv.org/abs/1505.04959}{{\tt 1505.04959}}.

\bibitem{Chu:1997ik}
C.-S. Chu and P.-M. Ho, ``{Poisson algebra of differential forms},'' {\em
  Int.J.Mod.Phys.} {\bf 12} (1997) 5573--5587,
\href{http://arXiv.org/abs/q-alg/9612031}{{\tt q-alg/9612031}}.

\bibitem{Beggs:2003ne}
E.~Beggs and S.~Majid, ``{Semiclassical differential structures},''
\href{http://arXiv.org/abs/math/0306273}{{\tt math/0306273}}.

\bibitem{McCurdy:2008ew}
A.~Tagliaferro, ``{The Star Product for Differential Forms on Symplectic
  Manifolds},''
\href{http://arXiv.org/abs/0809.4717}{{\tt 0809.4717}}.

\bibitem{McCurdy:2009xz}
S.~McCurdy and B.~Zumino, ``{Covariant Star Product for Exterior Differential
  Forms on Symplectic Manifolds},'' {\em AIP Conf.Proc.} {\bf 1200} (2010)
  204--214,
\href{http://arXiv.org/abs/0910.0459}{{\tt 0910.0459}}.

\bibitem{GenFCS} R.~Bonezzi, N.~Boulanger, E.~Sezgin and P.~Sundell,  
``{Generalized Frobenius-Chern-Simons gauge theories}'', in preparation.

\bibitem{Fujisawa:2013ima}
I.~Fujisawa, K.~Nakagawa, and R.~Nakayama, ``{AdS/CFT for 3D Higher-Spin
  Gravity Coupled to Matter Fields},'' {\em Class.Quant.Grav.} {\bf 31} (2014)
  065006,
\href{http://arXiv.org/abs/1311.4714}{{\tt 1311.4714}}.

\bibitem{Didenko:2008va}
V.~E. Didenko, A.~S. Matveev, and M.~A. Vasiliev, ``{Unfolded Description of
  $AdS_4$ Kerr Black Hole},'' {\em Phys. Lett.} {\bf B665} (2008) 284--293,
\href{http://arXiv.org/abs/0801.2213}{{\tt 0801.2213}}.

\bibitem{Vasiliev:2012vf}
M.~A. Vasiliev, ``{Holography, Unfolding and Higher-Spin Theory},'' {\em
  J.Phys.} {\bf A46} (2013) 214013,
\href{http://arXiv.org/abs/1203.5554}{{\tt 1203.5554}}.

\bibitem{Vasiliev:2015wma}
M.~A. Vasiliev, ``{Star-Product Functions in Higher-Spin Theory and
  Locality},'' {\em JHEP} {\bf 06} (2015) 031,
\href{http://arXiv.org/abs/1502.02271}{{\tt 1502.02271}}.

\bibitem{zachos2005quantum}
C.~Zachos, D.~Fairlie, and T.~Curtright, {\em Quantum mechanics in phase space:
  an overview with selected papers}, vol.~34.
\newblock World Scientific Publishing Company Incorporated, 2005.

\bibitem{Iazeolla:2007wt}
C.~Iazeolla, E.~Sezgin, and P.~Sundell, ``{Real Forms of Complex Higher Spin
  Field Equations and New Exact Solutions},'' {\em Nucl. Phys.} {\bf B791}
  (2008) 231--264,
\href{http://arXiv.org/abs/0706.2983}{{\tt 0706.2983}}.

\bibitem{Giombi:2013fka}
S.~Giombi and I.~R. Klebanov, ``{One Loop Tests of Higher Spin AdS/CFT},''
\href{http://arXiv.org/abs/1308.2337}{{\tt 1308.2337}}.

\bibitem{Kazinski:2005eb}
P.~O. Kazinski, S.~L. Lyakhovich, and A.~A. Sharapov, ``{Lagrange structure and
  quantization},'' {\em JHEP} {\bf 07} (2005) 076,
\href{http://arXiv.org/abs/hep-th/0506093}{{\tt hep-th/0506093}}.

\bibitem{Schwarz:1978cn}
A.~S. Schwarz, ``{The Partition Function of Degenerate Quadratic Functional and
  Ray-Singer Invariants},'' {\em Lett.Math.Phys.} {\bf 2} (1978)
247--252.

\bibitem{Schwarz:1984wk}
A.~S. Schwarz and Y.~Tyupkin, ``Quantization of antisymmetric tensors and
  ray-singer torsion,'' {\em Nucl.Phys.} {\bf B242} (1984)
436--446.

\bibitem{Horowitz:1989km}
G.~T. Horowitz and M.~Srednicki, ``{A Quantum Field Theoretic Description of
  Linking Numbers and Their Generalization},'' {\em Commun.Math.Phys.} {\bf
  130} (1990)
83--94.

\bibitem{Wu:1990ci}
S.-Y. Wu, ``{Topological Quantum Field Theories on Manifolds With a
  Boundary},'' {\em Commun.Math.Phys.} {\bf 136} (1991)
157--168.

\bibitem{Kontsevich:1997vb}
M.~Kontsevich, ``{Deformation quantization of Poisson manifolds. 1.},'' {\em
  Lett.Math.Phys.} {\bf 66} (2003) 157--216,
  \href{http://arXiv.org/abs/q-alg/9709040}{{\tt q-alg/9709040}}.

\bibitem{Cattaneo:1999fm}
A.~S. Cattaneo and G.~Felder, ``{A path integral approach to the Kontsevich
  quantization formula},'' {\em Commun. Math. Phys.} {\bf 212} (2000) 591--611,
\href{http://arXiv.org/abs/math/9902090}{{\tt math/9902090}}.

\bibitem{Gaberdiel:1997ia}
M.~R. Gaberdiel and B.~Zwiebach, ``{Tensor constructions of open string
  theories. 1: Foundations},'' {\em Nucl.Phys.} {\bf B505} (1997) 569--624,
\href{http://arXiv.org/abs/hep-th/9705038}{{\tt hep-th/9705038}}.

\end{thebibliography}
\end{document}